\pdfoutput=1
\documentclass[twocolumn,superscriptaddress,aps,preprintnumbers,amsmath,amssymb,prl,nofootinbib]{revtex4}

\usepackage{amsmath}
\usepackage{amssymb}
\usepackage{amsfonts}
\usepackage{graphicx}
\usepackage{xcolor}
\usepackage{xfrac}
\usepackage{comment}
\usepackage{pifont}
\usepackage{physics}
\usepackage{fourier}
\usepackage{hyperref}
\usepackage{bm}
\usepackage{enumitem}
\usepackage{xfrac}
\usepackage{tcolorbox}

\definecolor{rossoferrari}{HTML}{D9073D}
\definecolor{mediumblue}{HTML}{0000CD}
\definecolor{forestgreen}{HTML}{228B22}
\definecolor{desy_blue}{HTML}{009EE2}
\definecolor{desy_orange}{HTML}{FD8800}
\definecolor{light_pink}{rgb}{1,0.4,0.4}
\definecolor{light_blue}{rgb}{0.284602,0.317763,0.963947}
\hypersetup{setpagesize=false,bookmarksnumbered=true,bookmarksopen=true,%
colorlinks=true,linkcolor=light_blue,urlcolor=rossoferrari,citecolor=rossoferrari,linktocpage=false}

\begin{document}


\preprint{MS-TP-26-15}

\title{New gravitational-wave templates for metastable\smallskip\\cosmic strings: Loop breaking versus network collapse}

\author{Doa Hashemi Asl}
\email{doa.hashemi@uni-muenster.de}
\affiliation{Institute for Theoretical Physics, University of M\"unster, 48149 M\"unster, Germany}

\author{Kai Schmitz}
\email{kai.schmitz@uni-muenster.de}
\affiliation{Institute for Theoretical Physics, University of M\"unster, 48149 M\"unster, Germany}
\affiliation{Kavli IPMU (WPI), UTIAS, The University of Tokyo, Kashiwa, Chiba 277-8583, Japan}



\begin{abstract}
Metastable cosmic strings are a common prediction of grand unified theories and act as a source of a gravitational-wave background (GWB) that can explain the 2023 pulsar timing array (PTA) signal. In this paper, we revisit the GWB signal from metastable strings, emphasizing the need to carefully distinguish between two different time scales: (i) $t_{\rm LB}$, the time scale of loop breaking because of spontaneous monopole nucleation on closed string loops, and (ii) $t_{\rm NC}$, the time scale of network collapse when string segments attached to monopoles begin to enter the Hubble horizon. We discuss under which conditions these two time scales are similar or far apart from each other and illustrate the resulting consequences for the GWB signal. In doing so, we generalize the description of the GWB signal from metastable strings to a three-parameter model in terms of the string tension $G\mu$ and the time scales $t_{\rm LB}$ and $t_{\rm NC}$, which allows us to unify the modeling of standard metastable strings with what is known as quasi-stable strings. In the limit of a large $t_{\rm LB}/t_{\rm NC}$ ratio, we, moreover, derive a compact analytical expression for the predicted GWB spectrum in excellent agreement with numerical results in the literature.  We thus conclude that our new templates for the GWB spectrum from metastable strings can be readily used in the analysis of future PTA data sets.
\end{abstract}


\date{\today}
\maketitle


\noindent
\textbf{Introduction}\,---\,In cosmology, cosmic strings~\cite{Vilenkin:1984ib,Hindmarsh:1994re,Vilenkin:2000jqa} are the relativistic counterparts of magnetic vortices in type-II superconductors: thin tube-like defects in field configurations that can form in the early Universe via the Kibble--Zurek mechanism~\cite{Kibble:1976sj,Kibble:1980mv,Zurek:1985qw}, i.e., in consequence of spontaneous symmetry breaking during a cosmological phase transition. String formation notably requires the vacuum manifold after symmetry breaking to have a nontrivial first homotopy group, $\pi_1(\mathcal{M}) \not\cong \{\mathbb{1}\}$\,---\,a condition that is, e.g., satisfied when a $U(1)$ symmetry becomes spontaneously broken.

Cosmic strings are predicted by numerous extensions of the Standard Model; popular examples are, e.g., scenarios that describe the spontaneous breaking of $U(1)_{B-L}$~\cite{Buchmuller:2013lra,Buchmuller:2019gfy,Blasi:2020wpy}, the Abelian gauge group whose gauge charge corresponds to baryon-minus-lepton number. In fact, models giving rise to cosmic strings can be naturally embedded into the multi-step symmetry breaking chains that one typically encounters in grand unified theories (GUTs), which establishes an intriguing connection between the phenomenology of cosmic strings and the physics of grand unification~\cite{Jeannerot:2003qv,Dror:2019syi,Fu:2023mdu}.

In GUT scenarios, it frequently happens that the condition of nontrivial $\pi_1$ is fulfilled only for a specific symmetry breaking step, but not for the full symmetry breaking chain. Consider, e.g., the breaking pattern $G \rightarrow H \rightarrow K$ such that $\pi_1(H/K)  \not\cong \{\mathbb{1}\}$ but $\pi_1(G/K) \cong \{\mathbb{1}\}$. In this case, the second step will give rise to metastable strings~\cite{Vilenkin:1982hm,Preskill:1992ck,Buchmuller:2023aus} that are unstable against the spontaneous nucleation of defects associated with the first step. A minimal model realizing this idea (``model M1'') would be the two-step breaking $SU(2)_A \rightarrow U(1)_A \rightarrow \emptyset$ via a Higgs triplet and a Higgs doublet~\cite{Chitose:2023dam,Chitose:2024pmz}. A simple extension (``model M2'') would be the breaking chain $SU(2)_A \times U(1)_B \rightarrow U(1)_A \times U(1)_B \rightarrow U(1)_C$, again via a Higgs triplet and a Higgs doublet~\cite{Buchmuller:2021dtt,Chitose:2025cmt}. In both models, the first symmetry breaking step results in the production of monopoles and the second step in the production of metastable strings. If both steps occur one after the other in the hot thermal phase of the early Universe, the strings formed in the second step attach to the monopoles, leading to a rapidly decaying monopole--string network~\cite{Dunsky:2021tih}.

On the other hand, if cosmic inflation occurs during or after the monopole-producing phase transition, but before the string-producing phase transition, all monopoles from the first step will be heavily diluted and the metastable strings from the second step can be potentially long-lived. In this paper, we shall focus precisely on this scenario\,---\,the production of metastable strings after cosmic inflation has strongly diluted any abundance of pre-existing monopoles\,---\,which is particularly relevant in light of recent progress in the field of gravitational-wave (GW) astronomy. 

Indeed, since 2023, pulsar timing array (PTA) collaborations around the globe (CPTA~\cite{Xu:2023wog}, EPTA and InPTA~\cite{EPTA:2023fyk}, MPTA~\cite{Miles:2024seg}, NANOGrav~\cite{NANOGrav:2023gor}, and PPTA~\cite{Reardon:2023gzh}) have reported evidence for a stochastic GW background (GWB) at nanohertz frequencies. The significance of the signal is still below the $5\,\sigma$ discovery threshold; the NANOGrav 15-year (NG15) data~\cite{NANOGrav:2023hde}, e.g., contains evidence for a GWB signal at a level of $3\sim4\,\sigma$. Still, the prospect of an imminent $5\,\sigma$ discovery of a GWB has attracted much attention in recent years. In the literature, two possible origins of the signal are commonly discussed: (i) a cosmic population of inspiraling supermassive black-hole binaries~\cite{NANOGrav:2023hfp,DOrazio:2023rvl} or (ii) new particle physics in the early Universe~\cite{NANOGrav:2023hvm,EPTA:2023xxk}. The latter option is arguably more speculative, but also promises to open a new window to particle physics at very high energies and the cosmology of our Universe at very early times~\cite{Caprini:2018mtu}.

Among all new-physics scenarios that can explain the 2023 PTA signal, metastable cosmic strings represent a particularly attractive example~\cite{NANOGrav:2023hvm} (see Ref.~\cite{Buchmuller:2020lbh} for earlier work). The closed loops in the string network act as an efficient source of gravitational radiation that can be active over large parts of cosmic history~\cite{Vilenkin:1981bx,Vachaspati:1984gt,Damour:2001bk,Damour:2004kw}. As a consequence, a network of cosmic strings is expected to yield a GWB signal spanning over many orders of magnitude in frequency whose precise shape encodes information about the cosmic expansion rate and the evolution of the string network itself. In this sense, GWs from cosmic strings can serve as a logbook of the cosmic expansion history~\cite{Cui:2017ufi,Cui:2018rwi,Gouttenoire:2019kij,Gouttenoire:2019rtn}. Moreover, thanks to their broadband signal, they represent a major target for future GW observations across large parts of the GW frequency spectrum, including LISA~\cite{Auclair:2019wcv,Blanco-Pillado:2024aca,Dimitriou:2025bvq} and LVK~\cite{LIGOScientific:2025kry}. 

Topologically stable strings turn out to produce a GWB spectrum that is too flat in comparison to the 2023 PTA signal. That is, if described in terms of the GW energy-density power spectrum $\Omega_{\rm GW}$, the spectral index $n_{\rm GW} = d\log_{10}\Omega_{\rm GW}/d\log_{10}f$ in the PTA band is too close to zero in the case of stable strings~\cite{Esmyol:2025ket} (see Refs.~\cite{Ellis:2020ena,Blasi:2020mfx,Blanco-Pillado:2021ygr} for earlier work). Metastable strings offer a solution to this problem, since the metastability of the network readily results in a power-law suppression of the GWB spectrum when moving from higher to lower values on the frequency axis~\cite{Buchmuller:2021mbb} (see Ref.~\cite{Leblond:2009fq} for earlier work). This phenomenological advantage over stable strings, in combination with their important role in GUT model building, has led to a great deal of interest in metastable strings since summer 2023~\cite{Antusch:2023zjk,Lazarides:2023rqf,Afzal:2023cyp,Ahmed:2023pjl,Afzal:2023kqs,Lazarides:2024niy,Ahmed:2024iyd,Antusch:2024nqg,Datta:2024bqp,Maji:2024cwv,Pallis:2024joc,Ahmad:2025dds,Hu:2025sxv,Antusch:2025xrs,Maji:2025thf,Pallis:2025epn,Ingoldby:2025wcl}.

In view of its importance for the interpretation of current and future PTA data, we shall revisit the GWB signal from metastable strings in this paper and generalize its description with regard to one crucial aspect. The evolution of a metastable-string network is characterized by two different time scales that are absent in the case of stable strings: (i) the time scale of network collapse (NC), $t_{\rm NC}$, and (ii) the time scale of loop breaking (LB), $t_{\rm LB}$. Here, $t_{\rm NC}$ denotes the time when local observers begin to observe how finite string segments with monopoles attached to their ends start to enter their Hubble horizon. At times $t < t_{\rm NC}$, local observers cannot yet distinguish the metastable network of long strings inside their own Hubble patch from a stable-string network in the standard scaling regime~\cite{Albrecht:1984xv,Albrecht:1989mk,Ringeval:2005kr,Blanco-Pillado:2011egf}. Around $t_{\rm NC}$, however, the finite length of long-string segments becomes apparent. These segments then collapse under their own tension, which marks the end of the scaling regime and halts the formation of new closed loops from the network. Meanwhile, $t_{\rm LB}$ denotes the time scale on which closed loops break up, turning into small segments on their part, because of spontaneous monopole nucleation~\cite{Monin:2008mp}.

In earlier work, it was assumed that the same dynamics are responsible for the breaking of infinitely long strings on super-Hubble scales and the breaking of closed loops on sub-Hubble scales, namely, vacuum tunneling in a way that can be regarded as the magnetic counterpart of the electric Schwinger effect in quantum electrodynamics. Consequently, the modeling of the GWB spectrum from metastable strings thus far was based on the assumption (in fact, strict identification) that $t_{\rm LB} \equiv t_{\rm NC} \equiv t_S$, where $t_S$ denotes the end of the scaling regime. Recently, however, the authors of Ref.~\cite{Tranchedone:2026lav} pointed out several effects that may significantly expedite string breaking on super-Hubble scales, e.g., an enhanced breaking rate at finite temperature shortly after network formation or the presence of a small but nonnegligible abundance of thermally produced monopoles to which strings may attach on super-Hubble scales.

These observations motivate us to generalize the description of the GWB signal from metastable strings from a two-parameter to a three-parameter model: on top of the standard parameters $G\mu$ (the string tension, i.e., energy per length) and $\sqrt{\kappa}$ (the hierarchy parameter controlling the LB time scale $t_{\rm LB}$), we also treat the NC time scale $t_{\rm NC}$ as a free parameter. In explicit microscopic models, it may be possible to compute $t_{\rm NC}$ as a function of $\sqrt{\kappa}$; however, even in this case the two functions $t_{\rm NC}(\sqrt{\kappa})$ and $t_{\rm LB}(\sqrt{\kappa})$ do not have to automatically agree with each other.

The goal of our paper thus is to disentangle the dependence of the GWB spectrum on the two time scales $t_{\rm LB}$ and $t_{\rm NC}$ and to discuss the resulting GWB signal across the $G\mu$--$\sqrt{\kappa}$--$t_{\rm NC}$ parameter space. As we will show, this generalization results in a greater variety of possible GWB spectra that can be produced by metastable strings. Continuously increasing the hierarchy parameter $\sqrt{\kappa}$ from values around $\sqrt{\kappa} \sim 7\cdots 8$ to $\sqrt{\kappa} \geq 10$, we are notably able to smoothly interpolate between the spectra of standard metastable strings and the spectra of what is known as quasi-stable strings in the literature~\cite{Lazarides:2022jgr,Lazarides:2023ksx,Maji:2026nkz}. We will discuss how this interpolation is implemented at the numerical level, while the underlying assumptions about the cosmic history may differ from those in Refs.~\cite{Lazarides:2022jgr,Lazarides:2023ksx,Maji:2026nkz}.

The authors of Ref.~\cite{Lazarides:2023ksx} demonstrated that the GWB spectrum at large values of the hierarchy parameter, $\sqrt{\kappa} \ge 10$, yields a good explanation of the 2023 PTA signal. This observation is essential for GUT model building, as it opens up opportunities for constructing models in which the two relevant symmetry breaking steps are separated by an appreciable hierarchy, i.e., where the associated energy scales differ by a factor $\sqrt{\kappa} \sim 10\cdots 30$. In this paper, we will go beyond earlier work on quasi-stable strings and derive analytical expressions for the GWB spectrum in the large-$\sqrt{\kappa}$ limit. In doing so, we will build upon the analysis in Ref.~\cite{Schmitz:2024gds}, which describes the GWB spectra from stable strings ``for pedestrians'' (i.e., via purely analytical methods wherever possible). As a result, we will obtain a fully analytical understanding of the GWB spectra from metastable strings at $\sqrt{\kappa} \geq 10$ for the phenomenologically most interesting values of $t_{\rm NC}$. 


\medskip\noindent
\textbf{GWB signal}\,---\,We begin by reviewing the computation of the  GWB signal from stable and metastable strings based on the velocity-dependent one-scale (VOS) model for the evolution of the long-string network~\cite{Kibble:1984hp,Martins:1996jp,Martins:2000cs}. In both cases, the total GWB signal emitted by the population of closed string loops on sub-Hubble scales can be written as 
\begin{equation}
\label{eq:OGW}
h^2\Omega_{\rm GW}\left(f\right) = \frac{8\pi}{3\,(H_0/h)^2}\left(G\mu\right)^2 \sum_{k=1}^{k_{\rm max}} P_k \,\mathcal{I}_k\left(f\right) \,,
\end{equation}
where $\Omega_{\rm GW}$ is defined as the GW energy density per logarithmic frequency interval in units of the critical density for a spatially flat Universe. $H_0 = 100\,h\,\textrm{km}/\textrm{s}/\textrm{Mpc}$ is the present-day value of the Hubble rate. In the following, we will always express the GWB signal in terms of $h^2\Omega_{\rm GW}$, which is independent of the dimensionless Hubble constant $h$ and only depends on the constant factor $H_0/h \simeq 3.24 \times 10^{-18}\,\textrm{Hz}$.

The $k$-sum in Eq.~\eqref{eq:OGW} runs over the harmonic excitations of the string loops, which result in GW emission at the discrete frequencies $f_k = 2k/\ell$ given a loop length $\ell$. The sum starts at $k=1$, corresponding to the fundamental mode of loop oscillations, and extends up to $k_{\rm max}$, corresponding to excitations with wavelength $\lambda_k = \ell/(2k)$ of the order of the string width $\delta$. Because of the enormous hierarchy between $\ell$ and $\delta$, the largest $k$ value, $k_{\rm max} \sim \ell/(2\delta)$, can be easily as large as, say, $k_{\rm max} \sim 10^{35}$ (see, e.g., Eq.~(63) in Ref.~\cite{Schmitz:2024hxw}). Evaluating the sum in Eq.~\eqref{eq:OGW} up to such large $k$ values is, however, numerically infeasible. In practice, we will therefore set $k_{\rm max} = 10^5$, which is sufficient in order to guarantee numerical convergence of all GWB spectra presented below. 

The first factor in the $k$-sum, $P_k$, quantifies the GW power emitted by the $k$-th harmonic of a loop in units of $G\mu^2$,
\begin{equation}
\label{eq:PGW}
P_k = \frac{\Gamma}{H_{k_{\rm max}}^q}\frac{1}{k^q} \,.
\end{equation}
Here, $H_{k_{\rm max}}^q$, the $k_{\rm max}$-th generalized harmonic number of order $q$, ensures that the total GW power is normalized to $\Gamma$, 
\begin{equation}
H_{k_{\rm max}}^q = \sum_{k=1}^{k_{\rm max}} \frac{1}{k^q} \,, \qquad \sum_{k=1}^{k_{\rm max}} P_k = \Gamma \,, \qquad \frac{dE}{dt} = - \Gamma G\mu^2 \,.
\end{equation}
In our analysis, we will set $\Gamma = 50$, as suggested by the numerical simulations in Ref.~\cite{Blanco-Pillado:2017oxo}, and $q = \sfrac{4}{3}$, reflecting the assumption that the GW emission from the network is dominated by GW bursts from cusps on string loops~\cite{Vachaspati:1984gt,Damour:2001bk}.

The power spectrum in Eq.~\eqref{eq:PGW} receives corrections from gravitational backreaction, which was recently studied in Ref.~\cite{Wachter:2024aos}. Gravitational backreaction results in the smoothing of small-scale structure on loops and thus turns $P_k$ into a time-dependent function, or more precisely, a function that depends on the loop evaporation fraction. On the one hand, gravitational backreaction can lead to sizable deviations from a standard cusp-like power spectrum. On the other hand, it has been shown in Ref.~\cite{Wachter:2024zly} that the ensuing consequences for the GWB spectrum remain rather moderate, after all, with the size of the corrections because of gravitational backreaction never exceeding the $30\,\%$ level. In this paper, we will therefore neglect gravitational backreaction, mostly for simplicity and because the aspect that we would like to highlight\,---\,loop breaking versus network collapse\,---\,is independent of the question of gravitational backreaction. It would, however, be interesting (and supposedly straightforward) to refine the GWB spectra presented here by including gravitational backreaction in future work. 

The final ingredient in Eq.~\eqref{eq:OGW} is the factor $\mathcal{I}_k$, which integrates the number density of non-self-intersecting loops and a redshift factor over all possible GW emission times, 
\begin{equation}
\label{eq:Ikf}
\mathcal{I}_k\left(f\right) = \frac{2k}{f} \int_{t_{\rm ini}}^{t_0} dt \: \left(\frac{a(t)}{a_0}\right)^5 n\left(\frac{2k}{f}\frac{a(t)}{a_0},t\right) \,.
\end{equation}
Here, $t_{\rm ini}$ represents the moment during the scaling regime of the long-string network when GW emission, as opposed to particle emission, becomes the dominant energy loss channel for string loops~\cite{Gouttenoire:2019kij}. The value of $t_{\rm ini}$ controls the properties of the GWB spectrum at its high-frequency end~\cite{Servant:2023tua,Hu:2025sxv}. For large $t_{\rm ini}$ and small $G\mu$, one especially enters the regime of so-called low-scale cosmic strings~\cite{Schmitz:2024hxw,Schmitz:2025uxv}. In the present work, we are, however, interested in the properties of the GWB spectrum at its low-frequency end, where the metastability of cosmic strings can lead to a relevant modification of the GWB signal. For our purposes, the value of $t_{\rm ini}$ is thus irrelevant, and we simply set it to a tiny value, such that no spectral turnover at high frequencies appears in any of the GWB spectra presented below. Meanwhile, $t_0 \simeq 13.8\,\textrm{Gyr}$ denotes the current age of the Universe. $a$ is the Friedmann--Lema\^itre--Robertson--Walker (FLRW) scale factor, and $a_0$ is its current value; in our convention, $a_0=1$.

The properties of the loop population are encoded in $n(\ell,t)$, where $n(\ell,t)d\ell$ represents the number density of loops with length between $\ell$ and $\ell + d\ell$ at time $t$. In Eq.~\eqref{eq:Ikf}, the length argument of $n(\ell,t)$ is evaluated at $\ell = 2k/(f\,a_0/a(t))$, which is the length that a loop must have at the time of GW emission such that its $k$-th harmonic gives rise to a present-day (i.e., redshifted) GW frequency $f$. The loop number density of stable strings in the Nambu--Goto approximation can be computed in the VOS model~\cite{Sousa:2013aaa},
\begin{equation}
\label{eq:nlt}
n_{\rm stable}\left(\ell,t\right) = \mathcal{F}\,\frac{C_*\,\Theta\left(t-t_*\right)\Theta\left(t_*-t_{\rm ini}\right)}{\alpha_*\left(\alpha_* + \Gamma G\mu + \dot{\alpha}_*t_*\right)t_*^4} \left(\frac{a(t_*)}{a(t)}\right)^3 \,,
\end{equation}
where all quantities with a star index are evaluated at $t_*$, i.e., the time of formation of a loop that has length $\ell$ at time $t$.

We assume that all new loops in the network form with a characteristic length $\ell_*$ that corresponds to a fixed fraction $\alpha_L$ of the correlation length of the long-string network, $L$,
\begin{equation}
\ell_*  = \alpha_*t_* \,, \qquad \alpha_* = \alpha_L\,\xi(t_*) \,, \qquad L(t_*) = \xi(t_*)\,t_* \,.
\end{equation}
The time evolution of the dimensionless correlation length $\xi$, together with the root-mean-squared velocity $\bar{v}$ of the long strings in the network, is governed by the VOS equations, which we solve numerically in a standard FLRW background. During radiation domination, we obtain $\xi_r \simeq 0.271$ and $\bar{v}_r \simeq 0.662$. At the same time, the numerical simulation in Ref.~\cite{Blanco-Pillado:2013qja} demonstrated that the distribution of initial loop lengths is sharply peaked around $\alpha_* = \ell_*/t_* \sim 0.1$ during radiation domination. Correspondingly, we fix $\alpha_L$ at $\alpha_L \simeq 0.369$ at all times, such that $\alpha_L\,\xi_r \simeq 0.100$ during radiation domination. A loop that forms at time $t_*$ with length $\ell_* = \alpha_* t_*$ subsequently shrinks at a rate $d\ell/dt = - \Gamma G\mu$ because of GW emission. Its length hence evolves linearly,
\begin{equation}
\label{eq:ellt}
\ell = \ell_* - \Gamma G\mu \left(t-t_*\right) \,,
\end{equation}
which can be rearranged as an implicit equation for $t_*$,
\begin{equation}
\label{eq:tstar}
t_* = \frac{\ell + \Gamma G\mu\,t}{\alpha_* + \Gamma G\mu} = \frac{\ell + \Gamma G\mu\,t}{\alpha_L\,\xi(t_*) + \Gamma G\mu} \,.
\end{equation}
We solve this implicit equation for $t_*$ for all relevant values of $\ell$ and $t$, using again our numerical VOS solution for $\xi$. 

The prefactor $\mathcal{F} \sim 0.1$ in Eq.~\eqref{eq:nlt} is an efficiency factor that accounts for the error introduced by the approximation of a single initial loop length $\ell_*$ at each time $t_*$~\cite{Blanco-Pillado:2013qja}; for definiteness, we set it to $\mathcal{F} = 0.1$. Finally, the factor $C_*$ in Eq.~\eqref{eq:nlt} originates from the normalization of the loop production function in the VOS model and can be written as~\cite{Auclair:2019wcv}
\begin{equation}
\label{eq:Cstar}
C_* = C(t_*) \,, \qquad C(t) = \frac{\tilde{c}}{\gamma}\frac{\bar{v}(t)}{\xi^3(t)} \,,
\end{equation}
where the parameter $\tilde{c}$ measures how efficiently loops can be chopped off from the long-string network. $\tilde{c}$ needs to be measured in numerical simulations; based on Refs.~\cite{Martins:2000cs,Blanco-Pillado:2013qja,Blanco-Pillado:2011egf}, we set it to $\tilde{c} = 0.23$. The Lorentz factor $\gamma$ accounts for the fact that loops are created with a nonzero velocity that is eventually redshifted. Large loops with initial length $\ell_* = \alpha_* t_*$ are typically produced with an initial velocity of around $v \simeq 1/\sqrt{2}$~\cite{Bennett:1989yp}, such that $\gamma = (1-v^2)^{-1/2} \simeq \sqrt{2}$; for definiteness, we therefore set $\gamma = \sqrt{2}$ in our analysis.

Equations~\eqref{eq:OGW} to \eqref{eq:Cstar} represent the machinery that is necessary to compute the GWB spectrum from topologically stable Nambu--Goto strings. Based on these results, it is straightforward to obtain the corresponding GWB signal from metastable strings. As long as one is only interested in the GW emission from closed sub-Hubble loops, one simply has to replace the loop number density in Eq.~\eqref{eq:nlt} by a corresponding expression for metastable strings~\cite{Buchmuller:2021mbb,NANOGrav:2023hvm},
\begin{equation}
n_{\rm meta}\left(\ell,t\right) = n_{\rm stable}\left(\ell,t\right)\,\Theta\left(t_{\rm NC}-t_*\right)\,\mathcal{E}\left(\ell,t\right) \,,
\end{equation}
where the two new factors account for the effects of network collapse and loop breaking, respectively. The Heaviside function encodes the modeling assumption that at times $t > t_{\rm NC}$, no new loops can be created by the network, while the exponential function $\mathcal{E}$ describes the depletion of the loop number density because of monopole nucleation,
\begin{equation}
\mathcal{E}\left(\ell,t\right) = \exp\left[-\Gamma_d\,\mathcal{A}\left(\ell,t\right)\right] \,.
\end{equation}
Here, $\Gamma_d$ denotes the rate of spontaneous monopole nucleation (i.e., Schwinger-like vacuum tunneling at zero temperature) per unit string length. Based on $\Gamma_d$, we define the fundamental time scale relevant for loop breaking as
\begin{equation}
t_{\rm LB} = \frac{1}{\sqrt{\Gamma_d}} \,.
\end{equation}
Meanwhile, $\mathcal{A}$ denotes the total worldsheet area of a loop embedded in a flat spacetime background that is born at time $t_*$ and that has length $\ell$ at time $t$,
\begin{align}
\mathcal{A}\left(\ell,t\right) & = \int_{t_*}^t dt' \left[\ell + \Gamma G\mu\left(t-t'\right)\right] \\
& = \ell\left(t-t_*\right) + \frac{1}{2}\,\Gamma G\mu\left(t-t_*\right)^2 \,.
\label{eq:Alt}
\end{align}
Clearly, $\ell\left(t-t_*\right)$ is the naive worldsheet area that the loop would have if it simply had constant length $\ell$ at all times, and $\sfrac{1}{2}\,\Gamma G\mu\left(t-t_*\right)^2$ is a correction reflecting the fact that the loop length actually evolves from $\ell + \Gamma G\mu(t-t_*)$ at time $t_*$ to $\ell$ at time $t$. In summary, we can thus write
\begin{equation}
\label{eq:nmeta}
n_{\rm meta}\left(\ell,t\right) = n_{\rm stable}\left(\ell,t\right)\,\Theta\left(t_{\rm NC}-t_*\right)\,e^{-\mathcal{A}\left(\ell,t\right)/t_{\rm LB}^2} \,. 
\end{equation}

Equation~\eqref{eq:nmeta} is of central importance for our discussion in this paper. It illustrates that the two time scales $t_{\rm NC}$ and $t_{\rm LB}$ are logically distinct from each other. They only numerically agree with each other if one is willing to make further assumptions, e.g., that both the collapse of the network and the breaking of sub-Hubble loops are governed by Schwinger-like vacuum tunneling at zero temperature. 

Before we continue our discussion of the GWB spectrum from metastable strings, we mention in passing that closed loops may not be the only relevant source of GWs in the case of metastable strings. Indeed, string segments, bounded by monopoles on both ends, may also emit GWs~\cite{Leblond:2009fq,Buchmuller:2021mbb}. Such segments are produced by the network in two different ways. First, long strings attached to monopoles are nothing but finite string segments bounded by monopoles; these segments enter the Hubble horizon around $t_{\rm NC}$. Second, loop breaking produces string segments bounded by monopoles on sub-Hubble scales. In both cases, segments may also break again because of monopole nucleation. 

If the monopoles at the end of string segments carry unconfined flux, it is expected that the segments quickly collapse, releasing their energy mostly in the form of particle radiation, before the monopole on one end of the string and the antimonopole on the other end of the string annihilate each other. This situation is, e.g., realized in model M2 mentioned in the introduction. In the case of unconfined flux, the GW emission from string segments is thus negligible. In Ref.~\cite{NANOGrav:2023hvm}, this hypothesis is referred to as the \textsc{meta-l} model, where the suffix \textsc{l} indicates that only GW emission from loops is taken into account. In model M1, on the other hand, monopoles carry no unconfined flux, which opens the possibility of efficient GW emission from string segments. In Ref.~\cite{NANOGrav:2023hvm}, this hypothesis is referred to as the \textsc{meta-ls} model, where the suffix \textsc{ls} indicates that GW emission from both loops and segments is taken into account.

Recently, the efficiency of GW emission from string segments in the \textsc{meta-ls} scenario has been questioned in Ref.~\cite{Chitose:2025qyt}. The authors of this work argue that, as a string segment collapses, the monopoles on its end constantly collide with thermal fluctuations on the segment. These collisions cause the segment to dissipate its energy in the form of particle radiation and prevent it from entering the oscillatory regime considered in earlier work. As a result, the GW emission from segments is again negligible. These arguments are supported by the numerical simulations in Ref.~\cite{Dvali:2022vwh}, which simulate the behavior of string segments in model M1, confirming efficient particle radiation and the absence of an oscillatory regime. In view of these results, we will neglect the possibility of GW emission from string segments in this paper. At the same time, we call for more work on the dynamics of segments (especially going beyond the parametric regime considered in Ref.~\cite{Dvali:2022vwh}), in order to settle the case and firmly rule out any appreciable GW signal from string segments once and for all, or identify remaining loopholes.  


\medskip\noindent
\textbf{Parameter space}\,---\,In this paper, we advocate for a new perspective on metastable strings, according to which this scenario should be treated as a three-parameter model:

\smallskip
\ding{202}~The first model parameter is the dimensionless string tension, $G\mu$, which is the main parameter controlling the phenomenology of cosmic strings and basically the only free parameter in the case of stable strings. Here, $G = 1/(8\pi M_{\rm Pl}^2)$ is Newton's constant, with $M_{\rm Pl} \simeq 2.435 \times 10^{18}$ being the reduced Planck mass, and $\mu$ is the dimensionful string tension (i.e., energy per string length), which is controlled by the energy scale of the spontaneous symmetry breaking that gives rise to the string network, $\mu \sim 2\pi \eta^2$~\cite{Bogomolny:1975de}.

\smallskip
\ding{203}~The second model parameter is the time scale of loop breaking, $t_{\rm LB}$. Comparing $t_{\rm LB}^2$ to the worldsheet area $\mathcal{A}$ of a loop that has length $\ell$ at time $t$ determines the exponential suppression factor in the loop number density in Eq.~\eqref{eq:nmeta}. In concrete microscopic models, one may be interested in computing $t_{\rm LB}$ as a function of the underlying fundamental parameters. In the semiclassical limit of an infinitely thin string and point-like monopoles, one finds~\cite{Vilenkin:1982hm,Preskill:1992ck}
\begin{equation}
\label{eq:Gammad}
t_{\rm LB} = \frac{1}{\sqrt{\Gamma_d}} \,, \qquad \Gamma_d = \frac{\mu}{2\pi}\,e^{-\pi\kappa} \,, \qquad \kappa = \frac{m^2}{\mu} \,,
\end{equation}
with monopole mass $m$. Note that, in this expression, the assumption of point-like monopoles corresponds to a large scale separation, $m \gg \sqrt{\mu}$, i.e., a large value of $\kappa$.

Going beyond this simplified treatment, the authors of Ref.~\cite{Chitose:2023dam} recently evaluated the bounce action describing the monopole nucleation on strings, $S_B$, in model M1 as a function of all relevant model parameters. In the limit of a large $SU(2)_A$ gauge boson mass, their results reproduce the simple estimate in Eq.~\eqref{eq:Gammad}, $S_B^0 = \pi \kappa$, up to an $\mathcal{O}(1)$ factor. In fact, throughout the relevant parameter space, the authors of Ref.~\cite{Chitose:2023dam} only find mild corrections to the bounce action and never any drastic deviations. An analogous determination of $S_B$ in model M2 is still pending at present. 

A precise computation of $S_B$, and hence $\Gamma_d$ and $t_{\rm LB}$, is essential in order to map the PTA constraints on $t_{\rm LB}$ onto the underlying model parameter space. At the same time, we emphasize that, by construction, fitting the GWB signal from metastable strings to PTA data will always only constrain $t_{\rm LB}$ (alongside $G\mu$ and $t_{\rm NC}$); it will never constrain any of the underlying model parameters directly. From a phenomenological point of view, it would suffice to work with $t_{\rm LB}$ as one of the free fit parameters. Expressing $t_{\rm LB}$ in terms of other parameters (i.e., gauge and Higgs boson masses) only introduces an extra layer of parameter relations that cannot be resolved by the PTA data. Having said that, we will still proceed in the usual way and replace $t_{\rm LB}$ with the hierarchy parameter $\sqrt{\kappa}$ in our analysis, simply because it represents the standard quantity in the literature to quantify the LB time scale. In doing so, we must, however, keep in mind that the precise relation between $\sqrt{\kappa}$ and $t_{\rm LB}$ can receive corrections in explicit microscopic models. 


\smallskip
\ding{204}~The third model parameter is the time scale of network collapse, $t_{\rm NC}$. In most of the literature on metastable strings thus far, $t_{\rm NC}$ was identified with $t_{\rm LB}$, reflecting the assumption that string breaking on super- and sub-Hubble scales is caused by the same dynamics, namely, vacuum tunneling at zero temperature. Indeed, assuming that monopole nucleation on long strings also occurs at a rate $\Gamma_d$ per unit string length, one can estimate that long strings with monopoles on their ends begin to enter the Hubble horizon around
\begin{equation}
\Gamma_d \ell \sim H \,, \qquad \ell \sim H^{-1} \, \qquad H \sim \frac{1}{t} \,,
\end{equation}
which can be solved for the NC time scale,
\begin{equation}
\label{eq:vacuumtunneling}
\textrm{Vacuum tunneling:}\quad t_{\rm NC} \sim t_{\rm LB} = \left(\frac{1}{4G\mu}\right)^{1/2}\frac{e^{\pi\kappa/2}}{M_{\rm Pl}} \,.
\end{equation}

Recently, the authors of Ref.~\cite{Tranchedone:2026lav}, however, pointed out that $t_{\rm NC}$ may, in fact, be much smaller than $t_{\rm LB}$ for various reasons, which we shall now comment on in turn:

\smallskip
\textit{Finite-$T$ tunneling:} At high temperatures $T$, shortly after the formation of the network, the bounce action $S_B$ can receive sizable thermal corrections that temporarily result in more frequent monopole nucleation on strings. A necessary condition for such a thermal enhancement of $\Gamma_d$ is a temperature above the critical value $T_0 \simeq 1/2\sqrt{\mu/\kappa}$. This condition may or may not be satisfied in realistic models. In models of hybrid inflation, e.g., that produce a string network at the end of inflation during a so-called waterfall transition, it is easy to avoid temperatures above $T_0$~\cite{Buchmuller:2013lra,Buchmuller:2011mw,Buchmuller:2012wn}. The simple reason for this statement is that, in models of this type, the symmetric vacuum is initially stabilized by a coupling to the inflaton field rather than by thermal corrections. 

On the other hand, if indeed temperatures above $T_0$ should be reached during the early stages of the network evolution, the enhanced decay rate will result in more frequent monopole nucleation on super-Hubble scales. Long strings with monopoles on their end will then begin to enter the Hubble horizon at an earlier time, $t_{\rm NC} < t_{\rm LB}$,   
\begin{equation}
\label{eq:tNC1}
\textrm{Finite-$T$ tunneling:}\quad t_{\rm NC} \sim \left(\frac{45}{128\,\pi^4g_*}\right)^{1/2}\frac{e^{2\sqrt{2\pi\kappa}}}{M_{\rm Pl}} \,,
\end{equation}
where we used Eqs.~(15) and (38) in Ref.~\cite{Tranchedone:2026lav} and assumed a reheating temperature around the symmetry breaking scale of the string-producing phase transition, $T_{\rm rh} \sim \eta$. In comparison to the standard vacuum-tunneling case, $t_{\rm NC}$ now scales like $t_{\rm NC} \sim \exp(2\sqrt{2\pi\kappa})$ instead of $t_{\rm NC} \sim \exp(\pi\kappa/2)$. 

\smallskip
\textit{Finite-$T$ monopoles:} If monopoles should carry unconfined $U(1)$ flux, scattering processes in the plasma can result in a thermal monopole abundance, even if all pre-existing monopoles were previously strongly diluted by inflation~\cite{Preskill:1979zi,Turner:1982kh}. This condition is not satisfied in toy model M1, in which no unbroken $U(1)$ symmetry remains at low energies, but it is satisfied in toy model M2, which preserves a $U(1)_C$, and can be easily satisfied in realistic GUT models.

For a reheating temperature below the energy scale of the monopole-producing phase transition, thermal monopole production is Boltzmann-suppressed. It is thus only relevant at early times, when it causes the long strings in the network to attach to monopoles on super-Hubble scales. Similar to the case of an enhanced decay rate $\Gamma_d$, monopole production at finite temperature hence leads to super-Hubble segments that can be shorter than in the standard vacuum-tunneling scenario and which correspondingly enter the Hubble horizon at an earlier time, $t_{\rm NC} < t_{\rm LB}$,   
\begin{equation}
\label{eq:tNC2}
\textrm{Finite-$T$ monopoles:}\quad t_{\rm NC} \sim \left(\frac{45\,G\mu}{256\,\pi^7g_*\kappa^3}\right)^{1/2}\frac{e^{2\sqrt{2\pi\kappa}}}{M_{\rm Pl}} ,
\end{equation}
where we used Eqs.~(25) and (38) in Ref.~\cite{Tranchedone:2026lav} and assumed again a reheating temperature around the symmetry breaking scale of the string-producing phase transition, $T_{\rm rh} \sim \eta$. As before, we find that the parametric dependence of $t_{\rm NC}$ on $\sqrt{\kappa}$ changes from $t_{\rm NC} \sim \exp(\pi\kappa/2)$ to $t_{\rm NC} \sim \exp(2\sqrt{2\pi\kappa})$. 

\smallskip
\textit{Inflationary monopoles:} The authors of Ref.~\cite{Tranchedone:2026lav} also consider a third reason why $t_{\rm NC}$ might be suppressed compared to $t_{\rm LB}$: monopole production from large field fluctuations during inflation. As in the previous case, the rate of monopole production is exponentially suppressed in this scenario, assuming an inflationary Hubble rate below the energy scale of the monopole-producing phase transition. Nonetheless, the monopole abundance at the end of inflation can be sufficient to result in the formation of shorter segments than in the standard vacuum-tunneling scenario.

The authors of Ref.~\cite{Tranchedone:2026lav} assume that the Hubble rate, or more precisely the Gibbons--Hawking temperature during inflation, $T_{\rm GH}$, remains above the  energy scale of the string-producing phase transition, $T_{\rm GH} = H_{\rm inf}/(2\pi) > \eta$, such that strings only form after inflation. However, together with the upper limit on the tensor-to-scalar ratio from observations of the cosmic microwave background, $r \lesssim 0.034$~\cite{Balkenhol:2025wms}, this assumption implies an upper limit on the energy scale $\eta$,
\begin{equation}
\eta \lesssim 7.27 \times 10^{12}\,\textrm{GeV}\: \left(\frac{r}{0.034}\right)^{1/2} \,,
\end{equation}
which, together with $\mu \sim 2\pi\eta^2$, yields an upper limit of
\begin{equation}
\label{eq:Gmulimit}
G\mu \lesssim 2.23 \times 10^{-12}\: \left(\frac{r}{0.034}\right) \,.
\end{equation}

Such small values of the string tension are phenomenologically less interesting. Indeed, for $G\mu$ as small as in Eq.~\eqref{eq:Gmulimit},  neither stable nor metastable strings have a chance to explain the 2023 PTA signal. The authors of Ref.~\cite{Tranchedone:2026lav} ignore the upper limit Eq.~\eqref{eq:Gmulimit} and discuss the GWB spectrum from metastable strings for $H_{\rm inf}/(2\pi) = \eta$  and $G\mu = 10^{-7}$ regardless. In our analysis, we will, by contrast, omit the possibility of monopole production during inflation. 

\smallskip
In summary, we conclude that it is by far not guaranteed that the two time scales $t_{\rm NC}$ and $t_{\rm LB}$ must coincide with each other. The standard identification $t_{\rm NC} \sim t_{\rm LB}$ remains a viable option. But as illustrated by the above discussion, $t_{\rm NC}$ may also be significantly suppressed compared to $t_{\rm LB}$. At the same time, it is obvious that any attempt at estimating $t_{\rm NC}$ in a given microscopic model comes with huge uncertainties. The relations in Eqs.~\eqref{eq:tNC1} and \eqref{eq:tNC2} present two examples of such an estimate. But as argued above, both estimates depend on several assumptions about the underlying GUT model and cosmological history. From a phenomenological bottom-up perspective, it therefore seems most appropriate to treat $t_{\rm NC}$ and $t_{\rm LB}$ as two independent time scales\,---\,which is precisely what we advocate for in this paper. 


\begin{figure*}
\includegraphics[width=\textwidth]{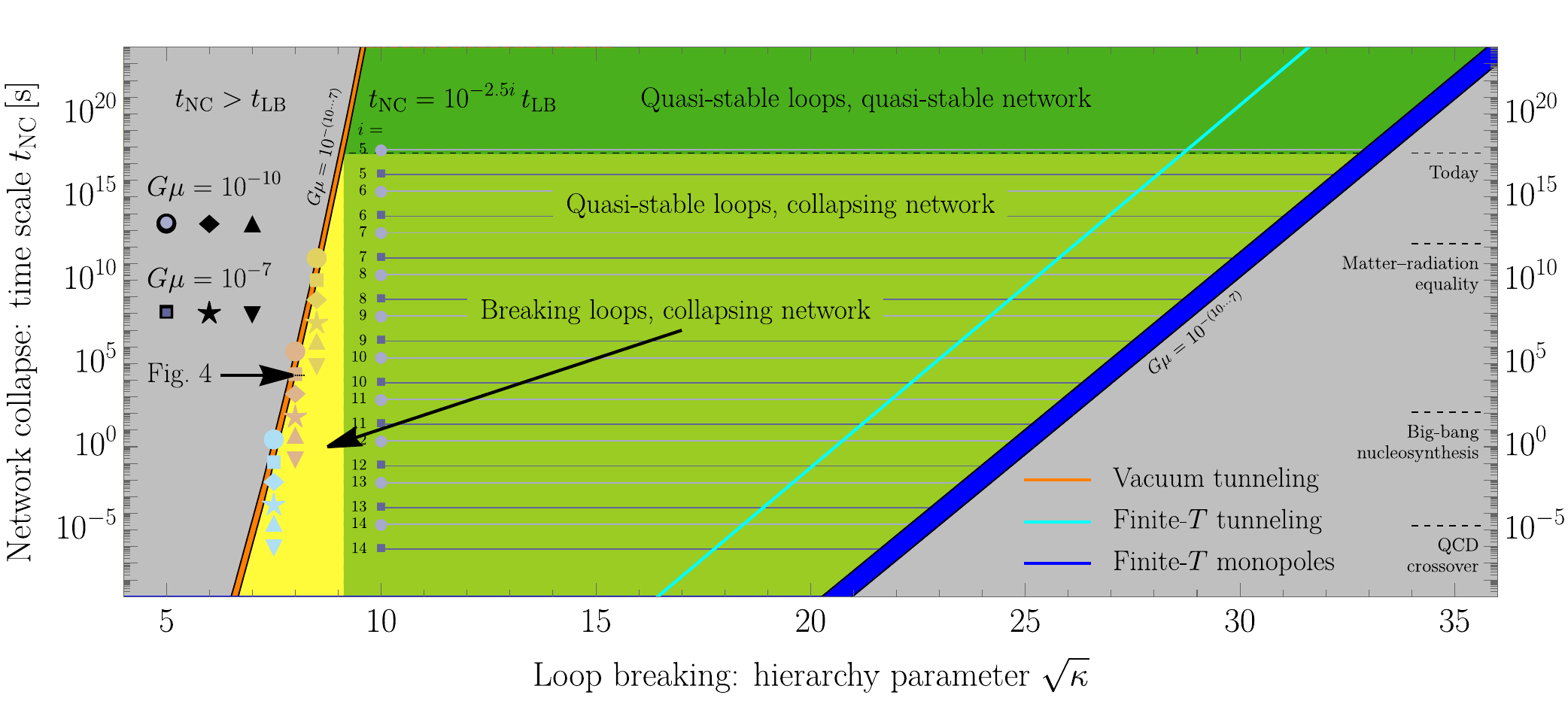}
\caption{Three regimes of metastable strings in the two-dimensional parameter plane spanned by $\sqrt{\kappa}$, which controls the time scale of loop breaking, $t_{\rm LB}$ [see Eq.~\eqref{eq:sqrtk}], and $t_{\rm NC}$, the time scale of network collapse: (i) $t_{\rm NC} < t_{\rm LB} < t_0$ (yellow shading), (ii) $t_0 < t_{\rm NC} < t_{\rm LB}$ (dark-green shading), and (iii) $t_{\rm NC} < t_0 < t_{\rm LB}$ (light-green shading). The orange line corresponds to the standard description of metastable strings [see Eq.~\eqref{eq:vacuumtunneling}], while the cyan and blue lines represent estimates of $t_{\rm NC}$ taken from Ref.~\cite{Tranchedone:2026lav} [see Eqs.~\eqref{eq:tNC1} and \eqref{eq:tNC2}]. The thickness of the orange and blue lines accounts for the dependence on the third relevant model parameter, the string tension $G\mu$. The various markers (circles, squares, diamonds, stars, and triangles) define benchmark points for which we show the respective GWB spectra in Figs.~\ref{fig:yellowspectra}, \ref{fig:quasi}, and \ref{fig:violins}.}
\label{fig:parameterspace}
\end{figure*}


\medskip\noindent
\textbf{Numerical results}\,---\,With the full machinery to compute the GWB signal from metastable strings at our disposal and with a clear understanding of the relevant parameters of the model, we are now ready to present our new GWB templates. The location of all GWB spectra that we will be discussing in the following is marked in Fig.~\ref{fig:parameterspace}, which depicts the $\sqrt{\kappa}$--$t_{\rm NC}$ parameter plane, while information on $G\mu$ is encoded in the different types of markers used in this plot. 

As illustrated in Fig.~\ref{fig:parameterspace}, we can distinguish three qualitatively different regions in the $\sqrt{\kappa}$--$t_{\rm NC}$ plane, depending on whether $t_{\rm LB}$ and $t_{\rm NC}$ are smaller or larger than the age of the Universe, $t_0$. Let us now discuss these three regimes in turn:

\smallskip
\textit{Breaking loops, collapsing network:} We define the first regime by demanding $t_{\rm LB}$ to remain smaller than $t_0$, which means that string loops are unstable against monopole nucleation and break apart at some point in the cosmic past. Meanwhile, $t_{\rm NC}$ is bounded from above by $t_{\rm LB}$. After all, if loop breaking is efficient on sub-Hubble scales, string breaking must also be efficient on Hubble and super-Hubble scales. The first regime, marked by the yellow shading in Fig.~\ref{fig:parameterspace}, is thus characterized by the hierarchy
\begin{equation}
t_{\rm NC} < t_{\rm LB} < t_0 \,. 
\end{equation}
In the limit $t_{\rm NC} \rightarrow t_{\rm LB}$, we notably recover the standard description of metastable strings that only depends on two parameters: $G\mu$ and the time scale $t_S = t_{\rm NC} = t_{\rm LB}$. In Fig.~\ref{fig:parameterspace}, this limiting case is indicated by the orange band, whose upper and lower edges correspond to $t_{\rm LB}$ as a function of $\sqrt{\kappa}$ evaluated for $G\mu = 10^{-10}$ and $G\mu = 10^{-7}$, respectively. The GWB spectra predicted by points along the orange band represent the standard GWB spectra from metastable strings in the literature~\cite{Buchmuller:2021mbb}, i.e., the spectra that were analyzed in Ref.~\cite{NANOGrav:2023hvm}. 

\begin{figure*}
\includegraphics[width=0.475\textwidth]{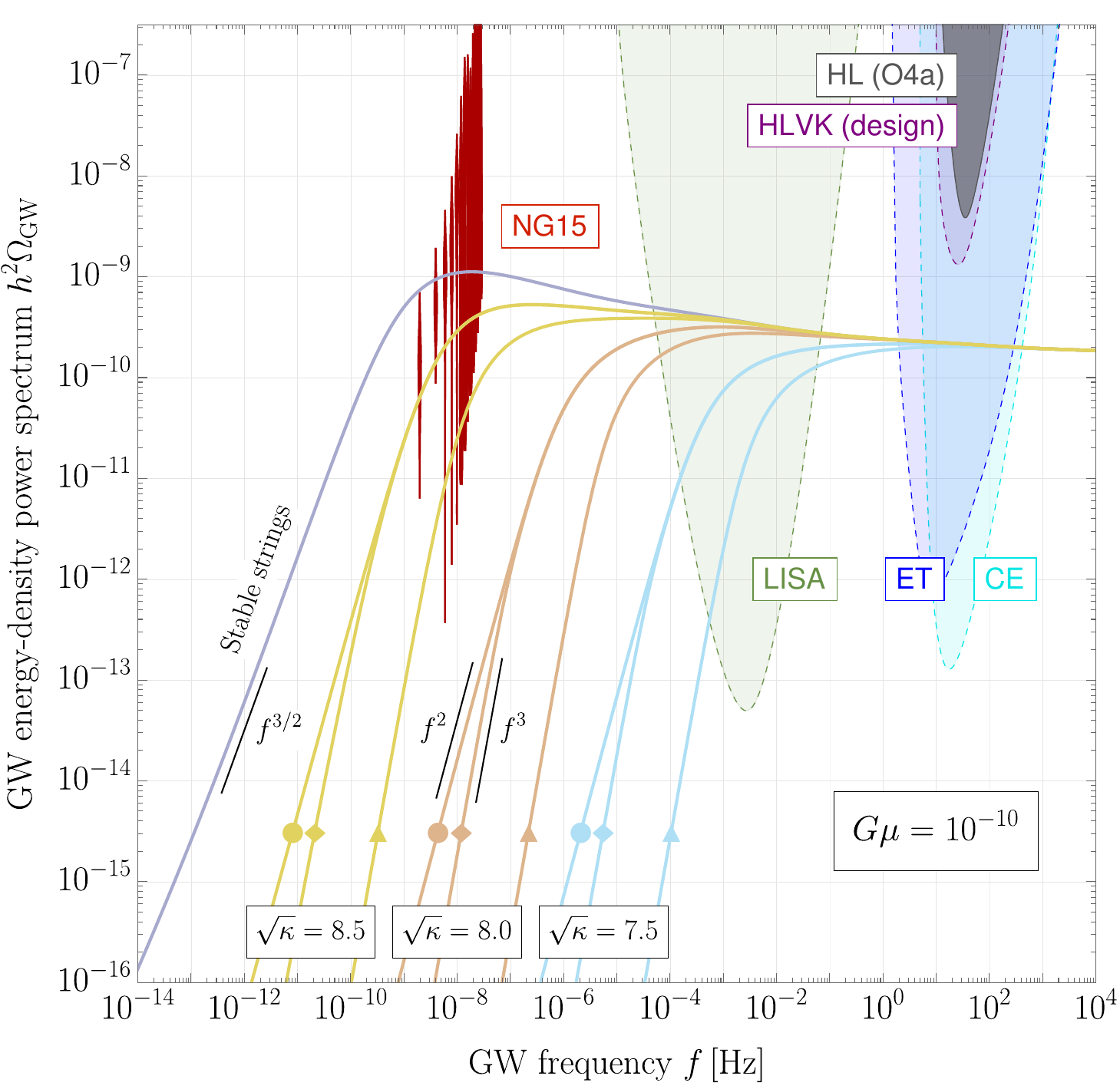}\quad
\includegraphics[width=0.475\textwidth]{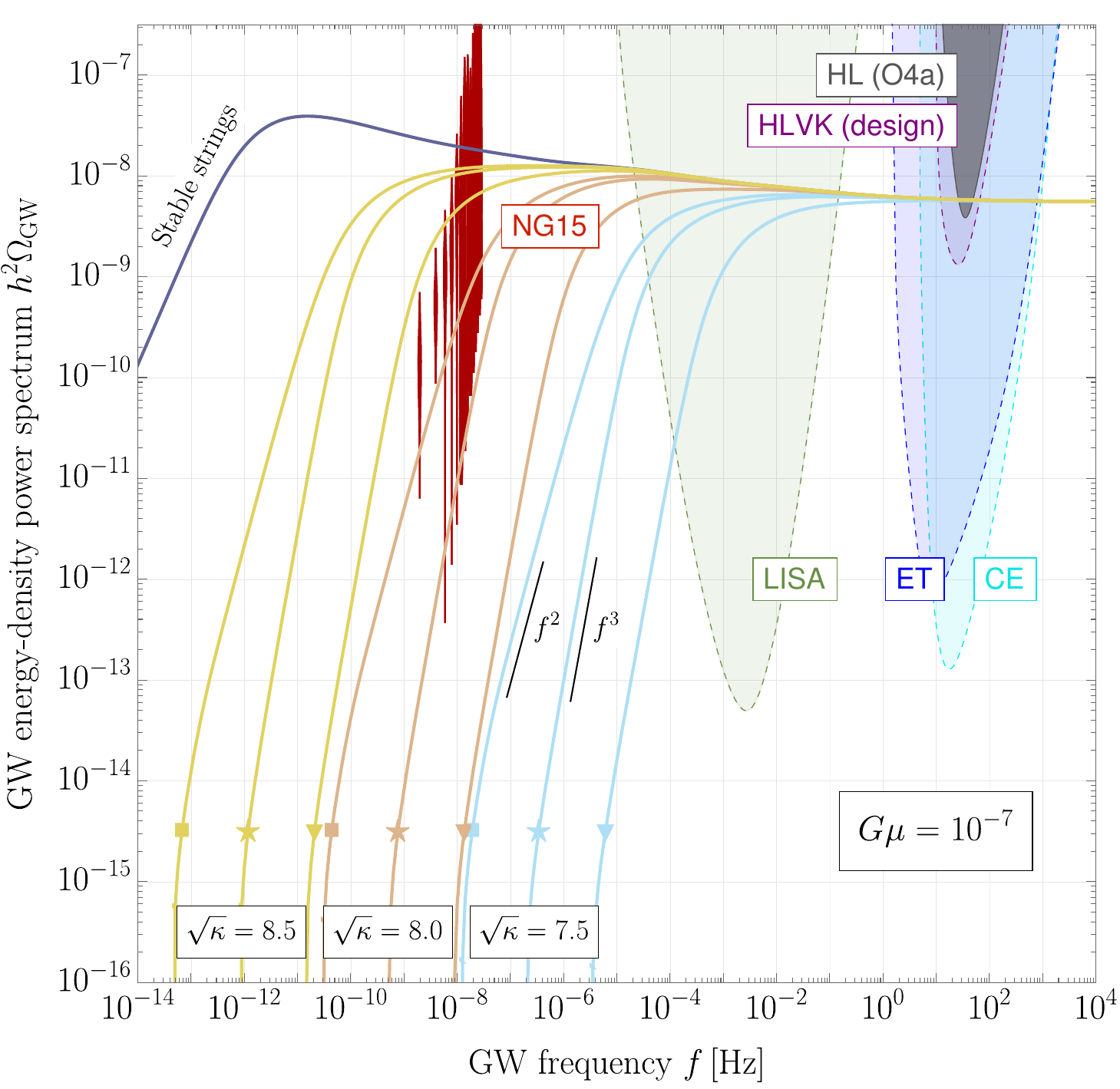}
\caption{GWB spectra for the benchmark points on the orange line (circles and squares) and inside the yellow-shaded region (diamonds, stars, and triangles) in Fig.~\ref{fig:parameterspace}. The former correspond to standard GWB spectra from metastable strings ($t_{\rm NC} = t_{\rm LB}$), while the latter represent a new parametric regime in which all three model parameters ($G\mu$, $\sqrt{\kappa}$, and $t_{\rm NC}$) matter and that has not been explored previously.}
\label{fig:yellowspectra}
\end{figure*}

If we now allow $t_{\rm NC}$ to be smaller than $t_{\rm LB}$, we enter new territory in the parameter space of metastable strings where the description of the resulting GWB spectrum genuinely relies on three parameters: $G\mu$, $\sqrt{\kappa}$, and $t_{\rm NC}$. To showcase the range of possible GWB spectra along the orange band and inside the yellow region in Fig.~\ref{fig:parameterspace}, we compute in total 18 benchmark spectra based on the machinery described above; see the two plots in Fig.~\ref{fig:yellowspectra}. Six of these spectra (marked by circles and squares) belong to the orange band, and twelve spectra (marked by diamonds, stars, and triangles) belong to parameter points inside the yellow region. The values of $G\mu$ and $\sqrt{\kappa}$ behind these spectra are indicated in Fig.~\ref{fig:yellowspectra}: $G\mu = 10^{-10}$ (left panel) and $G\mu = 10^{-7}$ (right panel) as well as $\sqrt{\kappa} = 7.5,8.0,8.5$ in both cases. Furthermore, for each combination of $G\mu$ and $\sqrt{\kappa}$, we consider three values of the NC time scale, $t_{\rm NC} = 10^{-5.0}\,t_{\rm LB}, 10^{-2.5}\,t_{\rm LB}, 10^{0.0}\,t_{\rm LB}$. 

In this region of parameter space, $\sqrt{\kappa}$ merely acts as a convenient parameter that allows us to parametrize the LB time scale in terms of an $\mathcal{O}(1)$ number. Indeed, we are free to ignore the fact that $\sqrt{\kappa}$ has originally been defined in terms of the monopole mass $m$ and string tension $\mu$ [see Eq.~\eqref{eq:Gammad}] and simply redefine $\sqrt{\kappa}$ in terms of $t_{\rm LB}$ [see Eq.~\eqref{eq:vacuumtunneling}],
\begin{equation}
\label{eq:sqrtk}
\sqrt{\kappa} = \left\{\frac{2}{\pi} \ln\left[\left(4G\mu\right)^{1/2}t_{\rm LB}\,M_{\rm Pl}\right]\right\}^{1/2} \,.
\end{equation}
With this definition of $\sqrt{\kappa}$ in mind, any PTA constraints on $\sqrt{\kappa}$ in the region of parameter space where $t_{\rm NC} < t_{\rm LB} < t_0 $ should then be understood as constraints on $t_{\rm LB}$ rather than constraints on the ratio of $m$ and $\sqrt{\mu}$. This change of perspective allows one to bypass the uncertainties in the computation of the bounce action $S_B$ that we discussed above: the PTA data allow us to directly constrain the LB time scale $t_{\rm LB}$, but they are not able to resolve further parameter relations that enter the computation of the bounce action $S_B$.

A full analytical description of the GWB spectra in this region of parameter space, following the ``pedestrian'' philosophy in Ref.~\cite{Schmitz:2024gds}, constitutes a nontrivial task because of the intricate interplay of the three parameters $G\mu$, $\sqrt{\kappa}$, and $t_{\rm NC}$. In the present paper, we will perform analytical calculations only for the limiting case in which the GWB spectrum ceases to depend on $\sqrt{\kappa}$ (see below); a comprehensive study of the GWB spectrum in the region of parameter space where all three parameters matter is left for future work~\cite{HashemiAsl:2026}.


For the purposes of the present paper, we content ourselves with the following observation: allowing for $t_{\rm NC} < t_{\rm LB}$ has nontrivial implications for the spectral index of the low-frequency tail of the GWB spectrum; see the ``power-law bars'' in Fig.~\ref{fig:yellowspectra}. To put this observation into perspective, we recall that earlier work on metastable strings first pointed to an $f^{3/2}$ power-law behavior at low frequencies~\cite{Buchmuller:2019gfy,Buchmuller:2020lbh}. A power law of this type follows from the GWB spectrum predicted by stable strings if one imposes a constant, frequency-independent cutoff in the lower integration boundary in Eq.~\eqref{eq:Ikf} deep in the radiation-dominated era~\cite{Schmitz:2024gds}, which reflects the approach in Refs.~\cite{Buchmuller:2019gfy,Buchmuller:2020lbh}.

This description was subsequently revised in Ref.~\cite{Buchmuller:2021mbb}, where the formalism outlined above was introduced. In this revised formulation, the lower integration boundary in Eq.~\eqref{eq:Ikf} can become frequency-dependent, which ultimately goes back to the $\ell(t-t_*)$ term in Eq.~\eqref{eq:Alt} and the fact that $\ell$ is set to $\ell = 2k/(f\,a_0/a(t))$ in the loop number density in Eq.~\eqref{eq:Ikf}. As shown in Ref.~\cite{Buchmuller:2021mbb}, this refined treatment of the lower integration boundary changes the power-law behavior at low frequencies from $f^{3/2}$ to $f^2$; see also Ref.~\cite{Dunsky:2021tih} for an independent analysis reaching the same conclusion.

Remarkably enough, we now find that the $f^2$ power-law behavior identified in Refs.~\cite{Buchmuller:2021mbb,Dunsky:2021tih} represents a somewhat tuned case that only holds true if $t_{\rm NC} \simeq t_{\rm LB}$. As soon as one allows for a hierarchy between these two time scales, $t_{\rm NC} \ll t_{\rm LB}$, the power-law behavior at low frequencies changes from $f^2$ to $f^3$. This observation corresponds to a new feature in the GWB spectrum from metastable strings that has not been noticed thus far and which may help improve the quality of the fit to the PTA signal. The point is the following: since the power law at low frequencies eventually turns over, approaching a nearly flat plateau at high frequencies, the spectral index needs to undergo a nontrivial evolution at intermediate frequencies. This evolution is sensitive to the details of the shape of the low-frequency spectrum, which may or may not improve the fit to the PTA signal.%
\footnote{In this paper, we restrict ourselves to presenting new templates for the GWB signal from metastable strings. As members of the NANOGrav collaboration, we will contribute to the application of these templates in official PTA fit analyses at the collaboration level in future work.}

Before we move on to the next regime in Fig.~\ref{fig:parameterspace}, we shall comment in passing on the sensitivity of the LIGO--Virgo--KAGRA (LVK) detector network to the GWB signal from metastable strings. In Fig.~\ref{fig:yellowspectra}, we indicate the LVK sensitivity reach in terms of power-law-integrated sensitivity (PLIS) curves~\cite{Thrane:2013oya} for the first part of the fourth observing run (O4a) as well as for the design sensitivity of the network. KAGRA joined the O4a run only for a few weeks in the beginning, and Virgo did not participate in the O4a run at all. In order to construct the LVK O4a PLIS curve, we therefore only work with the strain noise data of the LIGO detectors in Hanford (H) and Livingston (L) during the O4a run and neglect any Virgo (V) or KAGRA (K) contributions. Specifically, we employ the characteristic noise spectra shown in Fig.~5 of Ref.~\cite{Capote:2024rmo} and then compute the LVK O4a PLIS curve based on the formalism described in Ref.~\cite{Schmitz:2020syl} for a total observing time of $0.649\,\textrm{yr}$ (i.e., for the full duration of the O4a run). The PLIS curve thus obtained agrees very well with the ``$2\,\sigma$ PLIS curve'' shown in Fig.~7 of Ref.~\cite{LIGOScientific:2025bgj} if we normalize our PLIS curve to a signal-to-noise ratio (SNR) of $\rho = 3$. Consequently, we choose this SNR value for all PLIS curves shown in Fig.~\ref{fig:parameterspace}, including the curves for LISA (Laser Interferometer Space Antenna), CE (Cosmic Explorer), and ET (Einstein Telescope); for more details on PLIS curves, see Ref.~\cite{Schmitz:2020syl}.  

Comparing the LVK O4a PLIS curve with the GWB spectra for $G\mu = 10^{-7}$ in the right panel of Fig.~\ref{fig:yellowspectra}, we notice that a GWB signal from cosmic strings should have appeared in the O4a data with an SNR of $\rho >3$. However, the LVK collaboration has reported no such detection, which means that a string tension of $G\mu = 10^{-7}$ is (marginally) ruled out by the latest LVK data, whereas values around $G\mu \sim 10^{-7.5}$ are still in good shape. We do not attempt to make this upper limit more precise, as any such attempt would depend on further details, e.g., the choice of prior density for $G\mu$; see Ref.~\cite{Raidal:2026cpb} for a recent related discussion. Instead, we merely remark that our estimate of the maximal $G\mu$ value is consistent with the analysis by the LVK collaboration, which finds an upper limit around $G\mu \sim 10^{-(8\cdots6)}$ for a cosmic-string model that is roughly comparable to ours at LVK frequencies~\cite{LIGOScientific:2025kry} (since the LVK collaboration does not consider metastable strings, the respective string models differ from each other at lower frequencies). We thus conclude that a tension of $G\mu = 10^{-7}$ falls into the current LVK sensitivity reach, while tensions around $G\mu = 10^{-6}$ appear to be ruled out and tensions around $G\mu = 10^{-8}$ are still viable. 

\smallskip
\textit{Quasi-stable loops, quasi-stable network:} The second regime represents  the opposite of the previous case: NC and LB time scales much larger than the age of the Universe,
\begin{equation}
t_0 < t_{\rm NC} < t_{\rm LB} \,.
\end{equation}
In this parameter regime, marked by the dark-green shading in Fig.~\ref{fig:parameterspace}, the metastable-string network becomes indistinguishable from a stable-string network for local observers across all of cosmic history up to the present. In particular, $n_{\rm meta}$ in Eq.~\eqref{eq:nmeta} simply reduces to $n_{\rm stable}$ in this limit,
\begin{equation}
\lim_{t_{\rm NC},\,t_{\rm LB}\gg t_0} n_{\rm meta}\left(\ell,t\right) \rightarrow n_{\rm stable}\left(\ell,t\right) \quad \forall t \leq t_0 \,. 
\end{equation}

Correspondingly, the GWB signal from metastable strings in this parameter regime coincides with the standard GWB signal from stable strings. A comprehensive analytical discussion of this signal can be found in Ref.~\cite{Schmitz:2024gds}. In the present paper, we do not have anything new to add to this discussion. As far as the second regime is concerned, we content ourselves with including the GWB spectra for $G\mu = 10^{-10}$ and $G\mu = 10^{-7}$, which now no longer depend on $\sqrt{\kappa}$ and $t_{\rm NC}$, as reference spectra in the two panels of Fig.~\ref{fig:yellowspectra}.  

\smallskip
\textit{Quasi-stable loops, collapsing network:} The third regime assumes a position in the middle of the first two regimes. We now demand the NC and LB time scales to remain below and exceed the age of the Universe, respectively, 
\begin{equation}
t_{\rm NC} < t_0 < t_{\rm LB} \,, 
\end{equation}
which is realized in the region marked by the light-green shading in Fig.~\ref{fig:parameterspace}. The precise $\sqrt{\kappa}$ value at which $t_{\rm LB}$ begins to exceed $t_0$ depends on $G\mu$. However, as a rule of thumb, it suffices to note that, for all $G\mu$ values of interest, the condition $t_{\rm LB} > t_0$ is safely satisfied for $\sqrt{\kappa} \geq 10$; see Fig.~\ref{fig:parameterspace}. In the following, we shall therefore discuss the GWB spectrum from metastable strings for $\sqrt{\kappa} \geq 10$ and $t_{\rm NC} < t_0$. 

Before we turn to a more quantitative discussion, a few qualitative remarks are in order. First, we observe that, for $\sqrt{\kappa} \geq 10$, our cosmic-string model reduces again to a two-dimensional model: the remaining relevant parameters are $G\mu$ and $t_{\rm NC}$, and the loop number density $n_{\rm meta}$ reduces to
\begin{equation}
\label{eq:nmeta3}
\lim_{t_{\rm LB}\gg t_0} n_{\rm meta}\left(\ell,t\right) \rightarrow n_{\rm stable}\left(\ell,t\right)\,\Theta\left(t_{\rm NC}-t_*\right) \quad \forall t \leq t_0 \,. 
\end{equation}
In this sense, the parameter $\sqrt{\kappa}$ is no longer relevant in this region of parameter space. However, following the arguments in Ref.~\cite{Tranchedone:2026lav}, one may estimate $t_{\rm NC}$ as a function of $\sqrt{\kappa}$ in more specific cosmological scenarios; see Eqs.~\eqref{eq:tNC1} and \eqref{eq:tNC2}, which come with their own set of assumptions. In this case, $t_{\rm NC}$ can be traded for  $\sqrt{\kappa}$, and $\sqrt{\kappa}$ enters the analysis, after all. The discussion in Ref.~\cite{Tranchedone:2026lav} follows this logic, which alters the interpretation of the parameter $\sqrt{\kappa}$ once again: PTA constraints on $\sqrt{\kappa}$ in the parameter regime where $t_{\rm NC} < t_0 < t_{\rm LB}$ should not be interpreted as direct constraints on the ratio of $m$ and $\sqrt{\mu}$, but as constraints on $t_{\rm NC}$ via the relation that one has used to trade $t_{\rm NC}$ for $\sqrt{\kappa}$. 

In summary, this means the following: when $t_{\rm NC} < t_{\rm LB} < t_0$, one \textit{should} interpret $\sqrt{\kappa}$ as a measure for $t_{\rm LB}$, and when $t_{\rm NC} < t_0 < t_{\rm LB}$, one \textit{may} interpret $\sqrt{\kappa}$ as a measure for $t_{\rm NC}$. However, since this distinction may cause confusion, we do not advocate for a context-dependent interpretation of the parameter $\sqrt{\kappa}$ in this paper. Instead, we argue in favor of the following convention: $\sqrt{\kappa}$ simply allows for a convenient parametrization of $t_{\rm LB}$ according to Eq.~\eqref{eq:sqrtk}, and $t_{\rm NC}$ is an independent parameter that can be varied freely. In this paper, we consistently follow this convention\,---\,the only exception being Fig.~\ref{fig:parameterspace}, where we plot $t_{\rm NC}$ as a function of $\sqrt{\kappa}$ according to Eqs.~\eqref{eq:tNC1} and \eqref{eq:tNC2} for illustrative purposes. 

Next, let us turn to the GWB spectrum in the third parameter regime. Our analysis in this case will notably differ from the one in Ref.~\cite{Tranchedone:2026lav}, which implemented the following algorithm: first, consider a cosmological model, in which $t_{\rm NC}$ can be estimated in terms of $\sqrt{\kappa}$; then, identify the LB time scale with the NC time scale thus obtained; and finally, evaluate the standard GWB spectrum from metastable strings assuming $t_S = t_{\rm NC}(\sqrt{\kappa}) = t_{\rm LB}(\sqrt{\kappa})$. At the level of $n_{\rm meta}$, 
\begin{equation}
\label{eq:nmetatNC}
n_{\rm meta}\left(\ell,t\right) = n_{\rm stable}\left(\ell,t\right)\,\Theta\left(t_{\rm NC}-t_*\right)\,e^{-\mathcal{A}\left(\ell,t\right)/t_{\rm NC}^2} \,. 
\end{equation}
This form of $n_{\rm meta}$ reflects the physical assumption that whatever process results in more efficient string breaking on super-Hubble scales at early times is also responsible for more efficient loop breaking on sub-Hubble scales at late times. For instance, if finite-temperature effects should expedite the breaking of long strings at temperatures $T > T_0$, equation~\eqref{eq:nmetatNC} encodes the assumption that loop breaking will be characterized by a short LB time scale $t_{\rm LB} = t_{\rm NC}$ throughout the whole evolution of the network. We see no justification for this assumption: even if finite-temperature effects should affect the breaking of long strings on super-Hubble scales at early times, the breaking of string loops on sub-Hubble scales at late times will proceed via standard Schwinger-like vacuum tunneling at zero temperature. 

In the following, we will therefore refrain from working with Eq.~\eqref{eq:nmetatNC} and work with Eq.~\eqref{eq:nmeta3} instead. This approach leads to the realization that the GWB spectrum from metastable strings in the third parameter regime coincides with the GWB spectrum from what is known as quasi-stable strings in the literature~\cite{Lazarides:2022jgr,Lazarides:2023ksx,Maji:2026nkz}. Indeed, the analyses in Refs.~\cite{Lazarides:2022jgr,Lazarides:2023ksx,Maji:2026nkz} model the loop number density exactly according to Eq.~\eqref{eq:nmeta3}. The conclusion from this observation is the following: while the authors of Ref.~\cite{Tranchedone:2026lav} argue that the GWB spectrum from metastable strings does not change its spectral shape when allowing for parameter values such that $t_{\rm NC} < t_0 < t_{\rm LB}$ and that only the parametric mapping between spectral shapes and the underlying values of $\sqrt{\kappa}$ needs to be updated, we argue that the spectral shape is in fact different in the parameter region where $t_{\rm NC} < t_0 < t_{\rm LB}$. Rather, we argue that, in this limit, the signal from metastable strings coincides with the signal from what the authors of Refs.~\cite{Lazarides:2022jgr,Lazarides:2023ksx,Maji:2026nkz} refer to as quasi-stable strings. 

Meanwhile, we emphasize that our assumptions about the evolution of the metastable-string network are not identical to the assumptions in Refs.~\cite{Lazarides:2022jgr,Lazarides:2023ksx,Maji:2026nkz}. In a sense, the authors of Refs.~\cite{Lazarides:2022jgr,Lazarides:2023ksx,Maji:2026nkz} assume a mild tuning in their model, i.e., a temporal coincidence such that the monopole-producing phase transition only happens a few $e$-folds before the end of inflation. As a consequence, the primordial monopole abundance is less strongly diluted during inflation, and strings can attach more easily to monopoles during the string-producing phase transition. The outcome of the second phase transition is therefore a network of long string segments attached to monopoles that can enter the Hubble horizons of local observers at an early time $t_{\rm NC} \ll t_{\rm LB}$. Our framework is, by contrast, more model-agnostic and simply treats $t_{\rm NC}$ as a free input parameter from a bottom-up phenomenological perspective.

As we discussed, there may be many reasons why $t_{\rm NC} \ll t_{\rm LB}$: finite-temperature effects on super-Hubble scales shortly after the string-producing phase transition, thermal monopole production after inflation, and, indeed, a late monopole-producing phase transition during inflation. In this paper, we advocate for a parametrization of the GWB spectrum from metastable strings in terms of three parameters ($G\mu$, $\sqrt{\kappa}$, and $t_{\rm NC}$), which covers all of these cases at once and without the need to pick a particular model. 

\begin{figure}
\includegraphics[width=0.475\textwidth]{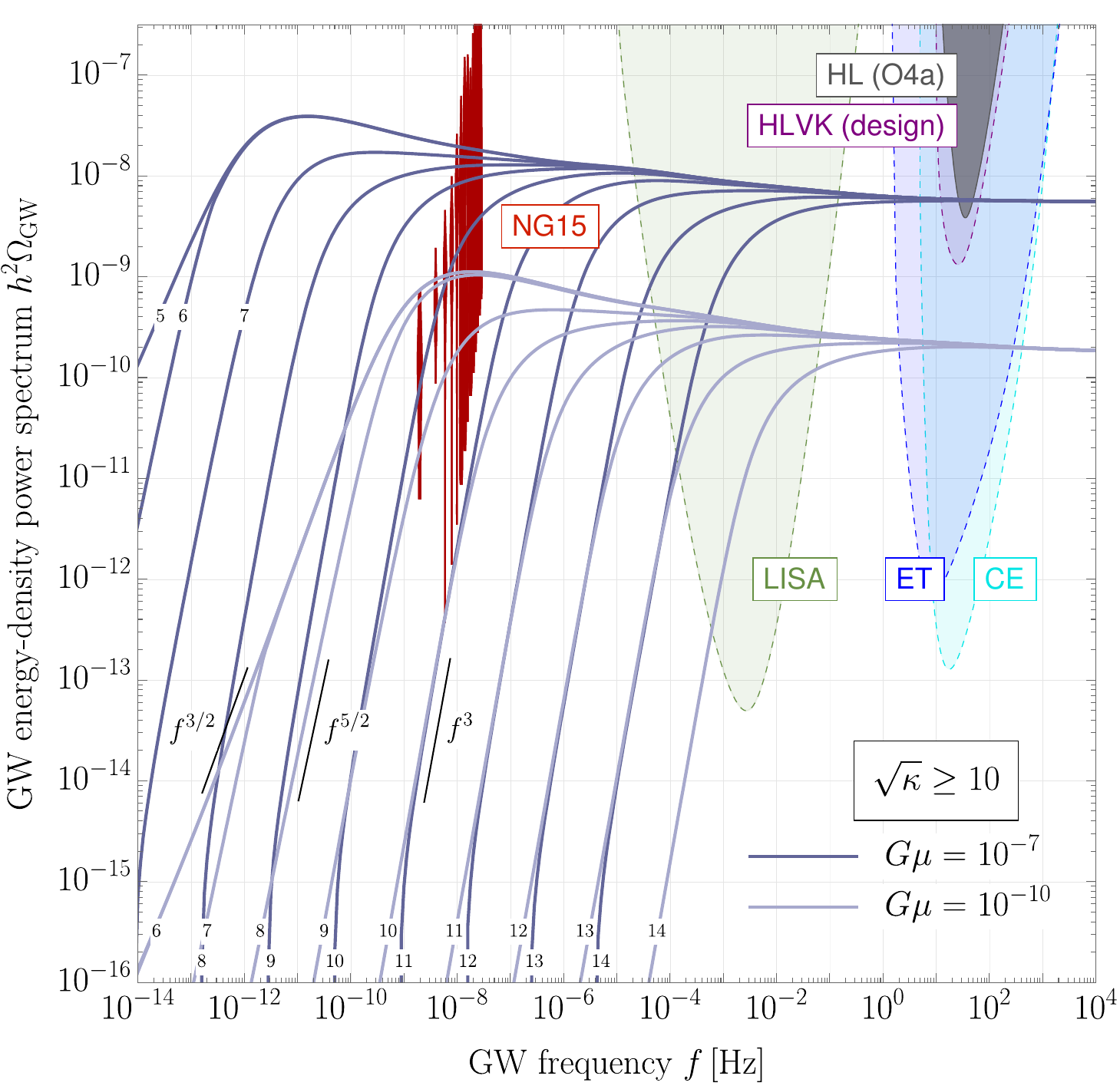}
\caption{GWB spectra for the benchmark points in the light-green region (circles and squares) in Fig.~\ref{fig:parameterspace}. These spectra fall into the class of spectra from quasi-stable strings~\cite{Lazarides:2022jgr,Lazarides:2023ksx,Maji:2026nkz} and only depend on $G\mu$ and $t_{\rm NC}$, but are independent of $\sqrt{\kappa}$. In Eq.~\eqref{eq:OmegaGWtot}, we provide a compact analytical expression describing these spectra in the parameter region that is relevant for the 2023 PTA signal.}
\label{fig:quasi}
\end{figure}

After these qualitative remarks, we are now ready to turn to our numerical results. In Fig.~\ref{fig:quasi}, we show the GWB spectra for the 20 benchmark points marked in the light-green region in Fig.~\ref{fig:parameterspace}. These points correspond again to $G\mu = 10^{-10}$ (spectra in light blue, circle markers in Fig.~\ref{fig:parameterspace}) and $G\mu = 10^{-7}$ (spectra in dark blue, square markers in Fig.~\ref{fig:parameterspace}). For definiteness, we set $\sqrt{\kappa} = 10$. However, since the GWB spectrum no longer depends on $\sqrt{\kappa}$ (in our convention) for all $\sqrt{\kappa} \geq 10$, our results automatically represent the predicted spectra across the entire light-green region in Fig.~\ref{fig:parameterspace}. In particular, this means that our results also describe the GWB spectra that one should expect when working with either of the expressions for $t_{\rm NC}$ in Eqs.~\eqref{eq:tNC1} and \eqref{eq:tNC2}. The second independent parameter in the light-green region is $t_{\rm NC}$, which we set to $t_{\rm NC} = 10^{-2.5i}t_{\rm LB}$, where $i=5,\cdots,14$, and $t_{\rm LB}$ is the $t_{\rm LB}$ value at $G\mu = 10^{-10}$ or $G\mu = 10^{-7}$ and $\sqrt{\kappa} = 10$; see Fig.~\ref{fig:parameterspace} for the exact location of these benchmark points. 

Interestingly enough, we now find three different types of power-law behavior at low frequencies: $f^{3/2}$, $f^{5/2}$, and $f^3$; see the ``power-law bars'' in Fig.~\ref{fig:quasi}. Meanwhile, we no longer find any $f^2$ behavior as in the case of the standard GWB spectrum from metastable strings (i.e., when $t_S= t_{\rm NC} = t_{\rm LB}$). We postpone a detailed discussion of these features again to future work. For the moment, we instead stress once more that the spectral shapes in the light-green region in parameter space do differ from the standard shapes. It is not possible to obtain the spectra in Fig.~\ref{fig:quasi} from the standard spectra for metastable strings and a simple parameter redefinition.


In order to illustrate this last point, we shall perform a final analysis before concluding our discussion. The benchmark points behind the spectra in Fig.~\ref{fig:quasi} correspond to a scan of parameter space in the vertical direction, i.e., along the $t_{\rm NC}$ axis. In addition, let us now scan the parameter space in the horizontal direction, i.e., along the $\sqrt{\kappa}$ axis. We start on the orange band at $G\mu = 10^{-7}$ and $\sqrt{\kappa} = 8.00$, where $t_{\rm NC} = t_{\rm LB} \simeq 19\,500\,\textrm{s}$, and then keep $t_{\rm NC}$ fixed at $t_{\rm NC}  \simeq 19\,500\,\textrm{s}$, while moving in small steps along the horizontal axis deeper into the yellow-shaded region. In total, we consider five benchmark points at $\sqrt{\kappa} = 8.00,\,8.05,\,8.10,\,8.15,\,8.20$; see the label, arrow, and five tiny points in Fig.~\ref{fig:parameterspace}.  

\begin{figure}
\includegraphics[width=0.475\textwidth]{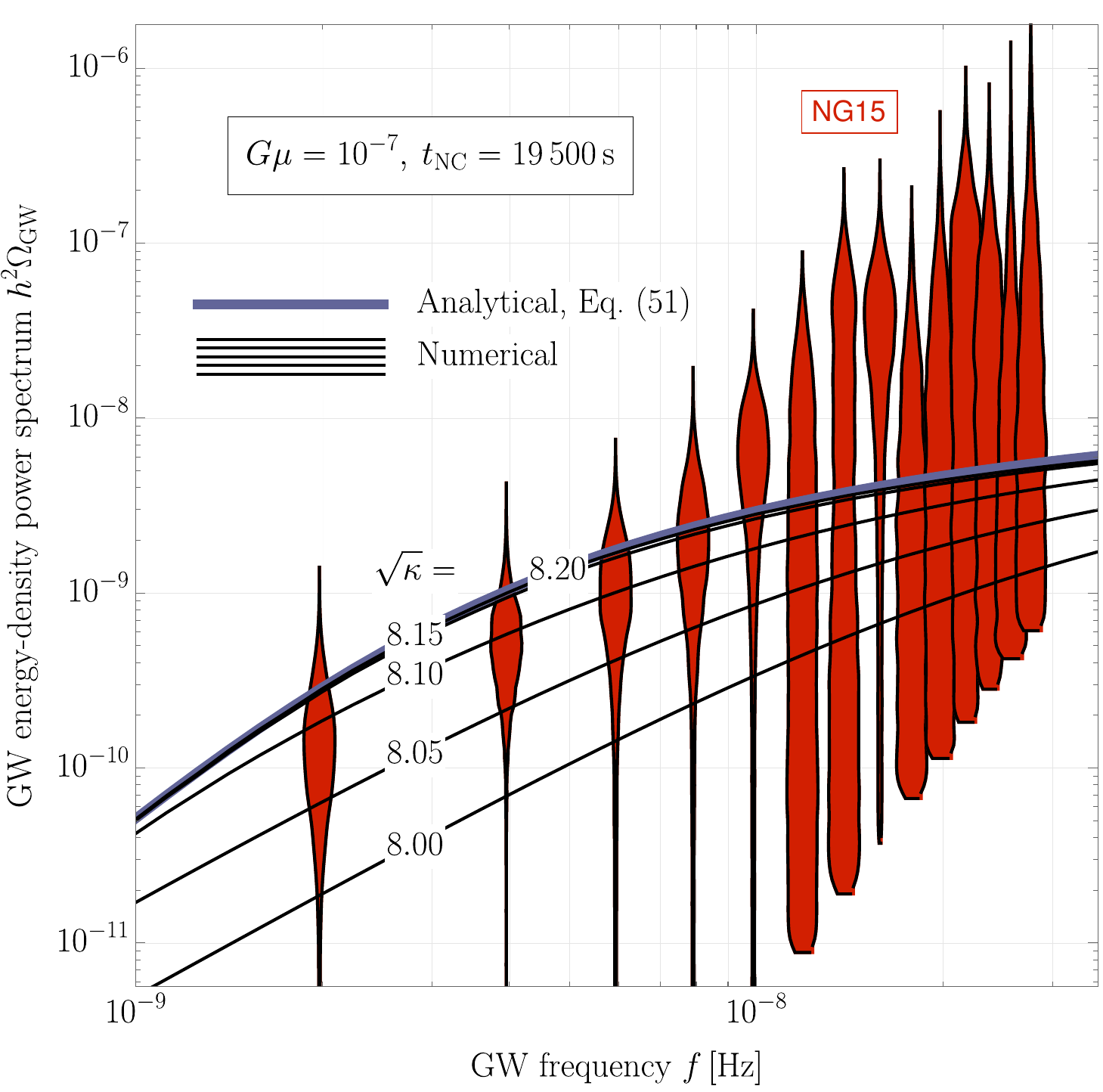}
\caption{GWB spectra for fixed $G\mu$ and $t_{\rm NC}$, but varying $\sqrt{\kappa}$; see the label, arrow, and five tiny points in Fig.~\ref{fig:parameterspace}. These spectra interpolate between a standard GWB spectrum from metastable strings and the asymptotic spectrum reached for large $\sqrt{\kappa}$ values. In contrast to Ref.~\cite{Tranchedone:2026lav}, we find that fixing $G\mu$ and $t_{\rm NC}$ does not uniquely determine the spectrum. The asymptotic spectrum falls into the class of spectra from quasi-stable strings~\cite{Lazarides:2022jgr,Lazarides:2023ksx,Maji:2026nkz}; in the PTA frequency range, it is described by the analytical expression in Eq.~\eqref{eq:OmegaGWtot}.}
\label{fig:violins}
\end{figure}

Following the arguments in Ref.~\cite{Tranchedone:2026lav}, all five benchmark points should predict the same GWB spectrum, simply because all five points share the same values of $G\mu$ and $t_{\rm NC}$. This is, however, not the case, as illustrated in Fig.~\ref{fig:violins}, where we show how the GWB spectrum changes its shape with increasing $\sqrt{\kappa}$, before it eventually does lose its sensitivity to $\sqrt{\kappa}$. The first spectrum ($\sqrt{\kappa} = 8.00$) belongs to a benchmark point on the orange band in Fig.~\ref{fig:parameterspace} and represents a standard GWB spectrum from metastable strings in the two-parameter regime, where $t_S= t_{\rm NC} = t_{\rm LB}$. The next three spectra ($\sqrt{\kappa} = 8.05,\,8.10,\,8.15$) lie in the yellow-shaded region in Fig.~\ref{fig:parameterspace} and represent spectra whose precise shape is sensitive to all three parameters: $G\mu$, $\sqrt{\kappa}$, and $t_{\rm NC}$. For these spectra, $t_{\rm LB}$ successively increases from $t_{\rm LB} \simeq 19.1\,\textrm{h}$ for $\sqrt{\kappa} = 8.05$ over $t_{\rm LB} \simeq 2.84\,\textrm{d}$ for $\sqrt{\kappa} = 8.10$ to $t_{\rm LB} \simeq 1.45\,\textrm{weeks}$ for $\sqrt{\kappa} = 8.15$. The last spectrum, finally, corresponds to $\sqrt{\kappa} = 8.20$, such that $t_{\rm LB} \simeq 5.24\,\textrm{weeks}$, exceeding $t_{\rm NC} \simeq 19\,500\,\textrm{s} \simeq 5.43\,\textrm{h}$ by more than a factor of $160$. We observe that, at such large values of the ratio $t_{\rm LB}/t_{\rm NC}$ and for the chosen fixed values of $G\mu$ and $t_{\rm NC}$, the spectrum begins to become insensitive to further increases in $\sqrt{\kappa}$.

In this sense, Fig.~\ref{fig:violins} vividly illustrates the transition from one type of spectral shape to another type of spectral shape. The starting point ($\sqrt{\kappa} = 8.00$) corresponds to a standard GWB spectrum from metastable strings based on the assumption that $t_S= t_{\rm NC} = t_{\rm LB}$, and the spectrum that is reached asymptotically for larger values ($\sqrt{\kappa} \gtrsim 8.20$) corresponds to a GWB spectrum from quasi-stable strings. In this way, Fig.~\ref{fig:violins} represents one of our main achievements in this paper: a unified description of the GWB signal from metastable and quasi-stable strings. In view of this unified description, we conclude that all scenarios discussed in this paper fall into the general category of metastable strings. Within this category, one may distinguish between the three different regimes discussed above. But we advise against a terminology in which metastable and quasi-stable strings represent two separate categories at the same level.


\medskip\noindent
\textbf{Analytical results}\,---\,The shape of the GWB spectrum for $\sqrt{\kappa} = 8.00$ in Fig.~\ref{fig:violins} can be understood based on the analytical discussion for standard metastable strings in Ref.~\cite{Buchmuller:2021mbb}. The spectra for $\sqrt{\kappa} = 8.05,\,8.10,\,8.15$, on the other hand, depend on all three parameters ($G\mu$, $\sqrt{\kappa}$, $t_{\rm NC}$), which is why we leave their analytical description for future work~\cite{HashemiAsl:2026}. Meanwhile, the spectrum for $\sqrt{\kappa} = 8.20$ can be reproduced analytically right away, as we are now going to demonstrate. In doing so, we will closely follow the analysis in Ref.~\cite{Schmitz:2024gds}. 

First, we note that the largest loops ever produced by the network are those that form at $t_* = t_{\rm NC}$ with initial length $\ell_* = \alpha_*t_{\rm NC}$. These loops are produced at the latest possible moment before the network collapses and require the longest time to evaporate because of GW emission. Neglecting loop breaking for a moment, we find from Eq.~\eqref{eq:ellt} that the largest loops in the network have shrunk to zero length at 
\begin{equation}
t_{\rm fin} = \left(\frac{\alpha_*}{\Gamma G\mu}+1\right)t_{\rm NC} \,.
\end{equation}
In combination with the fact that loops can break much earlier, $t_{\rm LB} \ll t_{\rm fin}$, the time $t_{\rm fin}$ represents an absolute upper limit on the lifetime of the entire network before all loops and segments have fully disappeared. For $G\mu = 10^{-7}$, $\alpha_* = 0.1$, $\Gamma = 50$, and $t_{\rm NC} \simeq 19\,500\,\textrm{s}$, we obtain $t_{\rm fin} \simeq 12.4\,\textrm{yr}$, which is still deep in the radiation-dominated era of the early Universe, $t_{\rm fin} \ll t_{\rm eq} \simeq 50\,400\,\textrm{yr}$. For this reason, we are able to restrict ourselves to loops that form and decay during radiation domination in the following, so-called RR loops~\cite{Schmitz:2024gds}. 

During radiation domination, the scale factor grows like $a(t) \propto t^{1/2}$. Plugging this relation into Eq.~\eqref{eq:nlt} and working with the solution for $t_*$ in Eq.~\eqref{eq:tstar} for constant $\alpha_* = \alpha_r = 0.1$, the number density of stable RR loops can be written as 
\begin{equation}
\label{eq:nstablerr}
n_{\rm stable}^{\rm rr}\left(\ell,t\right) = \mathcal{C}_r\,\frac{\Theta(t_{\rm eq}-t)\,\Theta(t-t_*)\,\Theta(t_*-t_{\rm ini})}{(\ell + \Gamma G\mu t)^{5/2}\,t^{3/2}} \,,
\end{equation}
where, for $\alpha_r \gg \Gamma G\mu$, the numerical prefactor evaluates to
\begin{equation}
\mathcal{C}_r = \mathcal{F}\,\frac{C_r\left(\alpha_r + \Gamma G\mu \right)^{3/2}}{\alpha_r} \approx \mathcal{F}\,C_r\,\alpha_r^{1/2} \simeq 0.171 \,.
\end{equation}
This expression for the loop number density, which we derived in the VOS model, is in excellent agreement with the loop number density that can be extracted from the numerical simulations by Blanco-Pillado, Olum, and Shlaer (BOS)~\cite{Blanco-Pillado:2013qja}. Indeed, the BOS expression exhibits the same functional form as our result, the only difference being that, instead of $\mathcal{C}_r \simeq 0.17$, BOS find a numerical prefactor of $0.18$. 


Based on Eq.~\eqref{eq:nstablerr}, let us now model the loop number density of metastable strings for large $\sqrt{\kappa}$ values as follows,
\begin{equation}
n_{\rm meta}\left(\ell,t\right) \approx n_{\rm stable}^{\rm rr}\left(\ell,t\right)\,\Theta\left(t_{\rm NC}-t_*\right) \,.
\end{equation}
This simple relation allows us to recycle much of the results for RR loops derived in Ref.~\cite{Schmitz:2024gds}; the presence of the additional Heaviside function merely tells us that we have to pay special attention to the integration boundaries when integrating over all possible GW emission times in Eq.~\eqref{eq:Ikf}.

The integral in Eq.~\eqref{eq:Ikf} assumes a particularly convenient form when we replace the time variable $t$ by a new variable
\begin{equation}
x_r = \frac{\Gamma G\mu}{4H_r^0}\frac{a_r(t)}{a_0}f \,,
\end{equation}
where $a_r(t)$ is the scale factor during radiation domination, and $H_r^0$ is a rescaled version of the Hubble constant, 
\begin{equation}
a_r\left(t\right) = a_0 \left(2H_r^0 t\right)^{1/2} \,, \qquad H_r^0 = H_0\,\Omega_r^{1/2} \,.
\end{equation}
In our analysis, we will only require $\omega_r = h^2\Omega_r$, i.e., the present-day value of the dimensionless density parameter for radiation, $\omega_r \simeq 4.18 \times 10^{-5}$~\cite{Fixsen:2009ug}. With these definitions, the GWB spectrum can be brought into the following form,
\begin{equation}
h^2\Omega_{\rm GW}\left(f\right) = \frac{1}{H_{k_{\rm max}}^q} \sum_{k=1}^{k_{\rm max}}\frac{1}{k^q}\,\Omega_{\rm GW}^{(1)}\left(\frac{f}{k}\right) \,, 
\end{equation}
where $\Omega_{\rm GW}^{(1)}$ corresponds to the GWB signal from loops oscillating in their fundamental oscillation mode, $k=1$, 
\begin{equation}
\label{eq:OmegaGW1}
h^2\Omega_{\rm GW}^{(1)}\left(f\right) = \mathcal{A}_r\,\Theta\left(x_r^{\rm end}-x_r^{\rm start}\right)\int_{x_r^{\rm start}}^{x_r^{\rm end}} dx_r\,\frac{\sfrac{3}{2}\,x_r^{1/2}}{\left(1+x_r\right)^{5/2}} \,. 
\end{equation}
The amplitude $\mathcal{A}_r$ in this expression is given by 
\begin{align}
\label{eq:Ar}
\mathcal{A}_r & = \frac{128\pi}{9}\,\mathcal{C}_r\omega_r\left(\frac{G\mu}{\Gamma}\right)^{1/2} \\
& \simeq 1.43 \times 10^{-8} \left(\frac{\mathcal{C}_r}{0.171}\right)\left(\frac{50}{\Gamma}\right)^{1/2}\left(\frac{G\mu}{10^{-7}}\right)^{1/2} \,,
\end{align}
and the lower boundary of the $x_r$ integral follows from requiring $t_{\rm ini} < t_* < t$, while the upper boundary of the $x_r$ integral follows from the conditions $t < t_{\rm eq}$ and $t_* < t_{\rm NC}$. 

It is now trivial to perform the $x_r$ integral in Eq.~\eqref{eq:OmegaGW1}, 
\begin{equation}
h^2\Omega_{\rm GW}^{(1)}\left(f\right) = \mathcal{A}_r\,\Theta\left(x_r^{\rm end}-x_r^{\rm start}\right)\,\left.\left(\frac{x_r}{1+x_r}\right)^{3/2}\right|_{x_r^{\rm start}}^{x_r^{\rm end}} \,. 
\end{equation}
For the parameters of interest, the two integration boundaries are separated by a large hierarchy. In the frequency range shown in Fig.~\ref{fig:violins}, the GWB spectrum is therefore only sensitive to the term induced by the upper boundary, 
\begin{equation}
\label{eq:OmegaGWxrend}
h^2\Omega_{\rm GW}^{(1)}\left(f\right) \approx \mathcal{A}_r\left(\frac{x_r^{\rm end}}{1+x_r^{\rm end}}\right)^{3/2}\,. 
\end{equation}

The only remaining task now is to determine the upper integration boundary. The condition $t < t_{\rm eq}$ translates to
\begin{equation}
\label{eq:xrend1}
x_r < x_r^{\rm eq} = \frac{\Gamma G\mu}{4H_r^0}\frac{a_{\rm eq}}{a_0}f \,,
\end{equation}
where $a_{\rm eq}$ represents the scale factor at matter--radiation equality, $a_0/a_{\rm eq} \simeq 3420$. Meanwhile, in order to determine the upper limit on $x_r$ that follows from the new condition $t_* < t_{\rm NC}$, we need to solve the quadratic equation
\begin{equation}
\label{eq:phi}
\varphi^2 + \varphi - \left(1+\chi_r\right)\left(x_r^{\rm NC}\right)^2 = 0 \,, 
\end{equation}
where $\varphi$ is a placeholder for $x_r^{\rm end}$, while $\chi_r = \alpha_r / (\Gamma G\mu)$ and 
\begin{equation}
x_r^{\rm NC} = \frac{\Gamma G\mu}{4H_r^0}\frac{a_r(t_{\rm NC})}{a_0}f \,.
\end{equation}
The quadratic equation in Eq.~\eqref{eq:phi} is solved by
\begin{equation}
\label{eq:phi2}
\varphi_2\left(x_r^{\rm NC},\chi_r\right) = \left[\left(1+\chi_r\right)\left(x_r^{\rm NC}\right)^2 + \frac{1}{4}\right]^{1/2} - \frac{1}{2} \,,
\end{equation}
which means that the condition $t_* < t_{\rm NC}$ can be written as
\begin{equation}
\label{eq:xrend2}
x_r < \varphi_2\left(x_r^{\rm NC},\chi_r\right) \,.
\end{equation}

For the benchmark point considered in Fig.~\ref{fig:violins}, $G\mu = 10^{-7}$, $\sqrt{\kappa} = 8.20$, and $t_{\rm NC} \simeq 19\,500\,\textrm{s}$, the upper limit in Eq.~\eqref{eq:xrend2} is more restrictive than the upper limit in Eq.~\eqref{eq:xrend1}, in accord with the fact that $t_{\rm NC} \ll t_{\rm eq}$. Introducing the frequency 
\begin{equation}
f_{\rm NC} = \frac{4H_r^0}{\Gamma G\mu}\frac{a_0}{a_r(t_{\rm NC})} = \frac{2}{\Gamma G\mu}\left(\frac{2H_r^0}{t_{\rm NC}}\right)^{1/2} \,,
\end{equation}
and, based on this quantity, the reference frequency
\begin{equation}
\label{eq:fstar}
f_* = \frac{f_{\rm NC}}{\left(1+\chi_r\right)^{1/2}} \approx \frac{f_{\rm NC}}{\chi_r^{1/2}} = \left(\frac{8H_r^0}{\alpha_r\Gamma G\mu\,t_{\rm NC}}\right)^{1/2} \,,
\end{equation}
we thus obtain for the upper integration boundary
\begin{equation}
\label{eq:xrend}
x_r^{\rm end} = \left[\left(\frac{f}{f_*}\right)^2 + \frac{1}{4}\right]^{1/2} - \frac{1}{2} \,,
\end{equation}
where $f_* \simeq 4.15\,\textrm{nHz}$ for our choice of parameters. 

Combining our results in Eqs.~\eqref{eq:OmegaGWxrend} and \eqref{eq:xrend}, we finally obtain the GWB signal from fundamental-mode oscillations, 
\begin{equation}
h^2\Omega_{\rm GW}^{(1)}\left(f\right) \approx \mathcal{A}_r\left(\frac{\left[\left(f/f_*\right)^2 + \sfrac{1}{4}\right]^{1/2} - \sfrac{1}{2}}{\left[\left(f/f_*\right)^2 + \sfrac{1}{4}\right]^{1/2} + \sfrac{1}{2}}\right)^{3/2}\,,
\end{equation}
and correspondingly the GWB signal from all $k$-modes,
\begin{equation}
\label{eq:OmegaGWtot}
h^2\Omega_{\rm GW}\left(f\right) \approx  \sum_{k=1}^{k_{\rm max}}\frac{\mathcal{A}_r'}{k^q}\left(\frac{\left[\left(f/(kf_*)\right)^2 + \sfrac{1}{4}\right]^{1/2} - \sfrac{1}{2}}{\left[\left(f/(kf_*)\right)^2 + \sfrac{1}{4}\right]^{1/2} + \sfrac{1}{2}}\right)^{3/2}\,,
\end{equation}
where we absorbed the generalized harmonic number in
\begin{equation}
\mathcal{A}_r' = \frac{\mathcal{A}_r}{H_{k_{\rm max}}^q}
\simeq 4.04 \times 10^{-9} \left(\frac{\mathcal{C}_r}{0.171}\right)\left(\frac{50}{\Gamma}\right)^{1/2}\left(\frac{G\mu}{10^{-7}}\right)^{1/2} \,.
\end{equation}

The expression in Eq.~\eqref{eq:OmegaGWtot} represents a second main achievement of our analysis in this paper. As illustrated in Fig.~\ref{fig:violins}, it yields an excellent approximation of the numerical GWB spectrum for $G\mu = 10^{-7}$, $\sqrt{\kappa} = 8.20$, and $t_{\rm NC} \simeq 19\,500\,\textrm{s}$. We thus succeeded in analytically solving the problem of quasi-stable strings, i.e., the problem of computing the GWB spectrum in regions of parameter space where the signal only depends on $G\mu$ and $t_{\rm NC}$, but no longer on $\sqrt{\kappa}$.

In Fig.~\ref{fig:violins}, we show our benchmark spectra in the PTA frequency band, alongside the NG15 free-spectrum ``violins'', in order to emphasize the relevance of our results for the interpretation of the PTA signal. As evident from Fig.~\ref{fig:violins}, the GWB spectra from metastable strings discussed in this paper offer an attractive explanation of the NG15 signal. The extra parametric freedom provided by the interplay of $G\mu$, $\sqrt{\kappa}$, and $t_{\rm NC}$ notably results in a greater variety of spectral shapes, which promises to improve the quality of the PTA fit beyond what is possible in the standard cases of metastable and quasi-stable strings present in the literature. 

Before we conclude, two more comments are in order. First, we emphasize that our result in Eq.~\eqref{eq:OmegaGWtot} involves the function $\varphi_2$ in Eq.~\eqref{eq:phi2}, which had already been identified in Ref.~\cite{Schmitz:2024gds}. As shown in Ref.~\cite{Schmitz:2024gds}, $\varphi_2$ also appears in the lower $x_r$ integration boundary (with a different $x_r$ value in the first argument), where it accounts for the condition $t_{\rm ini} < t_*$. The appearance of $\varphi_2$ in the lower $x_r$ integration boundary notably affects the shape of the GWB spectrum at high frequencies, where it turns over from a flat plateau to a declining power law. Earlier analytical work on the GWB signal from cosmic strings had overlooked this effect~\cite{Sousa:2020sxs}; see Fig.~3 in Ref.~\cite{Schmitz:2024gds}. In this sense, the discussion in Ref.~\cite{Schmitz:2024gds} had already pointed out the relevance of the function $\varphi_2$, albeit only in the high-frequency part of the spectrum, far away from the sensitivity reach of current GW experiments. A remarkable conclusion from our analysis now is that $\varphi_2$ can also affect other parts of the spectrum: specifically, the low-frequency part, when it appears in the upper integration boundary as a consequence of the condition $t_* < t_{\rm NC}$. This second appearance of $\varphi_2$ moves the spectral shape that it entails to a more accessible frequency range, the PTA band, thus defining a realistic observational target. 

The second comment in view of Eq.~\eqref{eq:OmegaGWtot} is that it yields a simple but powerful expression in the limit of low frequencies, $f\ll f_*$. Expanding up to leading order in $f/f_*$,  
\begin{align}
& h^2\Omega_{\rm GW}\left(f\right) \approx \mathcal{A}_r''\left(\frac{f}{f_*}\right)^3  \\
& \simeq 2.20 \times 10^{-11} \left(\frac{G\mu}{10^{-7}}\right)^2\left(\frac{t_{\rm NC}}{10^4\,\textrm{s}}\right)^{3/2}\left(\frac{f}{10^{-9}\,\textrm{Hz}}\right)^3 \,,
\label{eq:OGWlowf}
\end{align}
up to corrections of $\mathcal{O}((f/f_*)^5)$, and where we absorbed again a generalized harmonic number in the prefactor,
\begin{equation}
\mathcal{A}_r'' = H_{k_{\rm max}}^{q+3}\,\mathcal{A}_r' =  \frac{H_{k_{\rm max}}^{q+3}}{H_{k_{\rm max}}^q}\,\mathcal{A}_r \simeq 0.300\,\mathcal{A}_r \,.
\end{equation}
In Eq.~\eqref{eq:OGWlowf}, a factor of $(G\mu)^{1/2}$ originates from the $(G\mu)^{1/2}$ factor in $\mathcal{A}_r$ in Eq.~\eqref{eq:Ar}, while a factor of $(G\mu\,t_{\rm NC})^{3/2}$ follows from the $(G\mu\,t_{\rm NC})^{1/2}$ factor in $f_*$ in Eq.~\eqref{eq:fstar} raised to the third power. In light of the NG15 signal in Fig.~\ref{fig:violins}, we observe that GWB spectra similar to the one in Eq.~\eqref{eq:OGWlowf}\,---\,with an amplitude of around $2 \times 10^{-11}$ at a frequency of around $1\,\textrm{nHz}$\,---\,yield a good fit of the data. This requirement can be translated to a relation between the parameters $G\mu$ and $t_{\rm NC}$, 
\begin{equation}
\left(\frac{G\mu}{10^{-7}}\right)^2\left(\frac{t_{\rm NC}}{10^4\,\textrm{s}}\right)^{3/2} \sim 1 \,,
\end{equation}
which matches the outcome of the Bayesian fit analysis in Ref.~\cite{Lazarides:2023ksx}. Indeed, as shown in Ref.~\cite{Lazarides:2023ksx}, fitting quasi-stable strings to the NG15 data results in a strong anticorrelation in the two-dimensional posterior distribution for $G\mu$ and $t_{\rm NC}$ according to which $t_{\rm NC} \propto 1/(G\mu)^{4/3}$. Our analytical discussion above provides an explanation for this observation. 


\medskip\noindent
\textbf{Conclusions}\,---\,The GWB signal from metastable strings offers an attractive interpretation of the 2023 PTA signal. Thus far, the standard approach in modeling this signal relied on two parameters: the string tension $G\mu$ and the hierarchy parameter $\sqrt{\kappa}$, reflecting the underlying assumption that the breaking of sub-Hubble closed loops and the breaking of super-Hubble string segments are governed by the same dynamics: Schwinger-like vacuum tunneling at zero temperature. In this paper, following up on the analysis in Ref.~\cite{Tranchedone:2026lav}, we lifted this assumption and generalized the description of the GWB signal from metastable strings to a three-parameter model in terms of $G\mu$, $\sqrt{\kappa}$ (or $t_{\rm LB}$), and $t_{\rm NC}$. Here, $t_{\rm LB}$ is the time scale of loop breaking  because of spontaneous monopole nucleation on closed strings loops, and $t_{\rm NC}$ is the time scale of network collapse when segments attached to monopoles begin to enter the Hubble horizon.

We first described the general machinery allowing one to compute the GWB spectrum as a function of $G\mu$, $\sqrt{\kappa}$, and $t_{\rm NC}$; then, we gave a detailed discussion of the relevant parameter space and the three qualitatively different regimes in it; and finally, we computed and discussed the GWB spectra for a large number of representative benchmark points across parameter space. This analysis led us to several interesting results: (i) We managed to identify new spectral shapes and power-law behaviors that had not been described previously (see Fig.~\ref{fig:yellowspectra}); (ii) we unified the description of the GWB signals from metastable and quasi-stable strings~\cite{Lazarides:2022jgr,Lazarides:2023ksx,Maji:2026nkz} (see Figs.~\ref{fig:quasi} and \ref{fig:violins}); and (iii) we derived an analytical expression [see Eq.~\eqref{eq:OmegaGWtot}] describing the GWB spectrum at large $\sqrt{\kappa}$ values in the PTA frequency band. 

Remaining open questions pertain to the role of string segments and their possible contribution to the total GWB signal as well as to the reliability of the modeling assumptions that we made when writing down the loop number density in Eq.~\eqref{eq:nmeta}. On the one hand, more work is needed to better understand if string segments do indeed quickly lose all their energy via particle emission, or if they may be loopholes to this conclusion. On the other hand, the dynamics of metastable-string networks need to be studied in numerical lattice simulations, in order to improve the modeling of the abrupt decline in the production of new loops around the time of network collapse, $t \sim t_{\rm NC}$. In this paper, we described this abrupt change in the loop production by a Heaviside function; however, numerical lattice simulations may allow one to work out a smoother step function. Finally, it will be interesting to improve the computation of the bounce action $S_B$ in explicit microscopic cosmic-string models, in order to better match the parameters that are actually constrained by the PTA data ($G\mu$, $t_{\rm LB}$, and $t_{\rm NC}$) to the underlying parameters at the Lagrangian level.  

In summary, we conclude that our new templates for the GWB spectrum from metastable strings promise exciting applications for both theory and phenomenology. On the theory side, our templates support the conclusion of Ref.~\cite{Tranchedone:2026lav} that model-building efforts no longer have to be restricted to scenarios that predict a small hierarchy parameter, $\sqrt{\kappa} \lesssim 10$. Instead, the ratio of $m$ and $\sqrt{\mu}$ may be as large as $\mathcal{O}(30)$ (see Fig.~\ref{fig:parameterspace}), thus allowing for a larger hierarchy between the energy scales of the associated phase transitions. Furthermore, on the phenomenology side, our templates offer a broader variety of spectral shapes, which promises to lead to an even better fit of the PTA data. We plan to carry out such a PTA fit analysis in the near future. 


\medskip\noindent
\textbf{Acknowledgments}\,---\,D.\,H.\,A.\ acknowledges support from Studienstiftung des Deutschen Volkes. 
K.\,S.\ is an affiliate member of the Kavli Institute for the Physics and Mathematics of the Universe (Kavli IPMU) at the University of Tokyo and supported by the World Premier International Research Center Initiative (WPI), MEXT, Japan (Kavli IPMU).


\small
\bibliographystyle{utphys}
\bibliography{arxiv_1}


\end{document}